\pdfoutput=1 
\documentclass[%
 reprint,
superscriptaddress,
 amsmath,amssymb,
 aip,
]{revtex4-1}

\usepackage{graphicx}
\usepackage{dcolumn}
\usepackage{bm}
\usepackage{color, colortbl}
\usepackage{longtable}
\usepackage{multirow}
\usepackage{amsmath}
\usepackage[version=4]{mhchem}
\usepackage{enumitem}

\definecolor{Gray}{gray}{0.85}

\begin{document}


\title{Electronic Structure of Mononuclear Cu-based Molecule from Density-Functional Theory with Self-Interaction Correction}%

\author{Anri Karanovich}
\affiliation{Department of Physics, Virginia Tech, Blacksburg, Virginia 24061, USA}%

\author{Yoh Yamamoto}%
\affiliation{Department of Physics, University of Texas at El Paso, El Paso, Texas 79968, USA}

\author{Koblar Alan Jackson}
\affiliation{Physics Department and Science of Advanced Materials Program, Central Michigan University, Mt. Pleasant, Michigan 48859, USA}%

\author{Kyungwha Park}
\email{kyungwha@vt.edu (corresponding author)}
\affiliation{Department of Physics, Virginia Tech, Blacksburg, Virginia 24061, USA}%


\begin{abstract}
We investigate the electronic structure of a planar mononuclear Cu-based molecule [Cu(C$_6$H$_4$S$_2$)$_2$]$^z$ in two oxidation states
($z$$=$$-2$, $-$1) using density-functional theory (DFT) with Fermi-L\"owdin orbital (FLO) self-interaction correction (SIC). The dianionic Cu-based molecule was proposed to be a promising qubit candidate. Self-interaction error within approximate DFT functionals renders severe delocalization of electron and spin densities arising from 3$d$ orbitals. The FLO-SIC method relies on optimization of Fermi-L\"owdin orbital descriptors (FODs) with which localized occupied orbitals are constructed to create the SIC potentials. Starting with many initial sets of FODs, we employ a frozen-density loop algorithm within the FLO-SIC method to study the Cu-based molecule. We find that the electronic structure of the
molecule remains unchanged despite somewhat different final FOD configurations. In the dianionic state (spin $S=1/2$), FLO-SIC spin density originates from the Cu $d$ and S $p$ orbitals with an approximate ratio of 2:1, in quantitative agreement with multireference calculations, while in the case of SIC-free DFT, the orbital ratio is reversed. Overall, FLO-SIC lowers the energies of the occupied orbitals and in particular the 3$d$ orbitals unhybridized with the ligands significantly, which substantially increases the energy gap between the highest occupied molecular orbital (HOMO) and 
the lowest unoccupied molecular orbital (LUMO) compared to SIC-free DFT results. The FLO-SIC HOMO-LUMO gap of the dianionic state is larger than 
that of the monoionic state, which is consistent with experiment. Our results suggest a positive outlook of the FLO-SIC method in the description
of magnetic exchange coupling within 3$d$-element based systems.
\end{abstract}

\date{\today}

\maketitle

\section{Introduction} \label{sec:intro}


Systems including 3$d$ transition-metal elements are difficult to study using density-functional theory (DFT) due to strong electron correlation involving the localized $d$ orbitals. The approximate nature of the exchange-correlation functional within the DFT formalism limits an accurate
description of multiconfigurational/multireference features of strongly correlated systems. In addition, such nature allows significant Coulomb interactions of electrons with themselves, referred to as self-interaction error (SIE) \cite{Perdew1981}, which results in an underestimate of the band gap or gap between the highest occupied molecular orbital (HOMO) and the lowest unoccupied molecular orbital (LUMO) as well as an overestimate of exchange interaction between 3$d$ transition-metal centers, to name a few. An introduction of on-site Coulomb repulsion $U$ within the DFT formalism \cite{Anisimov1997,Kulik2006} aligns with an effort to compensate for SIE. Although standard multireference quantum chemistry methods can describe mononuclear transition-metal-based molecules, they are not practical for multinuclear magnetic molecules due to extremely high computational cost.

As an alternative, one can consider an application of DFT with self-interaction correction (SIC) to 3$d$ transition-metal systems. Perdew and Zunger (PZ) proposed a systematic method to impose the SIC to any spin-density-functional approximation \cite{Perdew1981}. This PZ-SIC formalism was successfully applied within the local spin density approximation (LSDA) to atoms and molecules \cite{Perdew1981,Pederson1984} and solids \cite{Heaton1983}. However, the PZ-SIC method adds a set of $N^2$ conditions that need to be satisfied to reach a minimum energy and the SIC energy is not necessarily size-consistent, where $N$ is the total number of electrons. Recently, a new practical SIC scheme adapted from the PZ-SIC formalism has been proposed using localized Fermi-L\"owdin orbitals (FLO) \cite{Pederson2014,Pederson2015}. In this scheme referred to as FLO-SIC, $3N$ conditions have to be satisfied and the SIC energy is size-consistent. The FLO-SIC method was applied to various non-magnetic molecules \cite{Pederson2016,Adhikari2020,Vargas2020,Jackson2019,Johnson2019,Joshi2018,Schwalbe2018}, giving rise to improvement of ionization energies of organic molecules \cite{Adhikari2020} and vertical detachment energies of water clusters \cite{Vargas2020}. However, applications of the FLO-SIC method to 3$d$ transition-metal systems have been limited \cite{Kao2017,Joshi2018,Jackson2019}. Small mononuclear 3$d$ transition-metal-based molecules would be ideal to gauge an applicability of the FLO-SIC method, the results of which provide insight into employment of the FLO-SIC method to multinuclear 3$d$ transition-metal systems.


Recently, a crystal of small Cu-based magnetic molecules, [Cu(C$_6$H$_4$S$_2$)$_2$]$^{-2}$ (or [Cu(II)(bdt$^{2-}$)$_2$]$^{-2}$), has been shown to have long spin-lattice and spin-spin relaxation times at room temperature attributed to small spin-orbit coupling, to the well-separated ground doublet, and to strong metal-ligand covalency \cite{Fataftah2019}, which renders the molecule a promising candidate for quantum information science applications. The electronic structure of the dianionic Cu-based molecule was investigated using electron paramagnetic resonance (EPR) experiments \cite{Fataftah2019,Maiti2014}, and it was also calculated using multireference quantum chemistry methods \cite{Fataftah2019}, time-dependent DFT (TDDFT) \cite{Maiti2014}, and DFT with a hybrid functional \cite{Maiti2014}. Interestingly, Ref.~\onlinecite{Maiti2014} suggested that two competing electronic structures (where the spin density is mostly carried by either the Cu or the S ligands) coexist in the dianionic Cu-based molecule, although such a scenario was not reported in the multireference study \cite{Fataftah2019}. On the other hand, the monoanionic Cu-based molecule, [Cu(C$_6$H$_4$S$_2$)$_2$]$^{1-}$, was observed to be non-magnetic with its UV-visible absorption peak at somewhat lower energy than the dianionic molecule \cite{Maiti2014}. Derivatives of the monoanionic Cu-based molecule were also synthesized and their properties including optical absorption spectra were characterized \cite{Ray2005,Robertson2002}. The electronic structures of the monoanionic molecule and one of the derivatives, [Cu(III)(C$_{14}$H$_{20}$S$_2$)$_2$]$^{1-}$, were calculated using DFT with a hybrid functional \cite{Maiti2014,Ray2005}.

In this work, we systematically investigate the electronic structure of the dianionic and monoanionic mononuclear Cu-based molecules,
[Cu(C$_6$H$_4$S$_2$)$_2$]$^{z}$ ($z=-2,-1$), using the FLO-SIC method. Starting with many Fermi-L\"owdin orbital descriptor (FOD) configurations
for a given molecular structure and charge state, we optimize them in order to obtain the optimal FLO-SIC energy for each FLO configuration.
For comparison, we also do SIC-free (standard) DFT and self-interaction free wave function calculations at the unrestricted Hartree-Fock (UHF)
and multiconfigurational levels. Since the FLO-SIC potential lowers energies of the occupied Cu $d$ orbitals much more than the occupied ligand orbitals, the FLO-SIC HOMO energy is greatly shifted downward which results in a large HOMO-LUMO gap. In addition, characteristics of the FLO-SIC 
HOMO significantly differ from those of the standard DFT and multiconfigurational/multireference calculations. The FLO-SIC spin density for the dianonic case quantitatively agrees well with the multiconfigurational/multireference result and it is consistent with EPR experimental data \cite{Maiti2014}. The FLO-SIC HOMO-LUMO gap of the monoanionic molecule is significantly smaller than that of the dianionic molecule, which 
may be in line with the experimental UV-optical absorption spectra.

In Sec.~\ref{sec:geo} we present the geometries of the dianionic and monoanionic Cu-based molecules that we use. In Sec.~\ref{sec:methods} we discuss our systematic applications of the FLO-SIC method to the molecules using a frozen-density loop algorithm, molecular symmetries and previous FLO-SIC results from atoms \cite{DY_Kao}. Then we show the FLO-SIC charge and spin density of states as well as the HOMO and LUMO characteristics for the two charge states compared to our UHF, standard (SIC-free) DFT, and multiconfigurational calculations as well as other groups' studies \cite{Fataftah2019,Maiti2014,Ray2005} in Secs.~\ref{sec:results-Q2} and \ref{sec:results-Q1}. In Sec.~\ref{sec:summary} we make conclusions and
provide outlook.




\section{Geometries of Cu-based molecules} \label{sec:geo}

\begin{figure}[h!]
    \centering
    \includegraphics[width=1.0\linewidth]{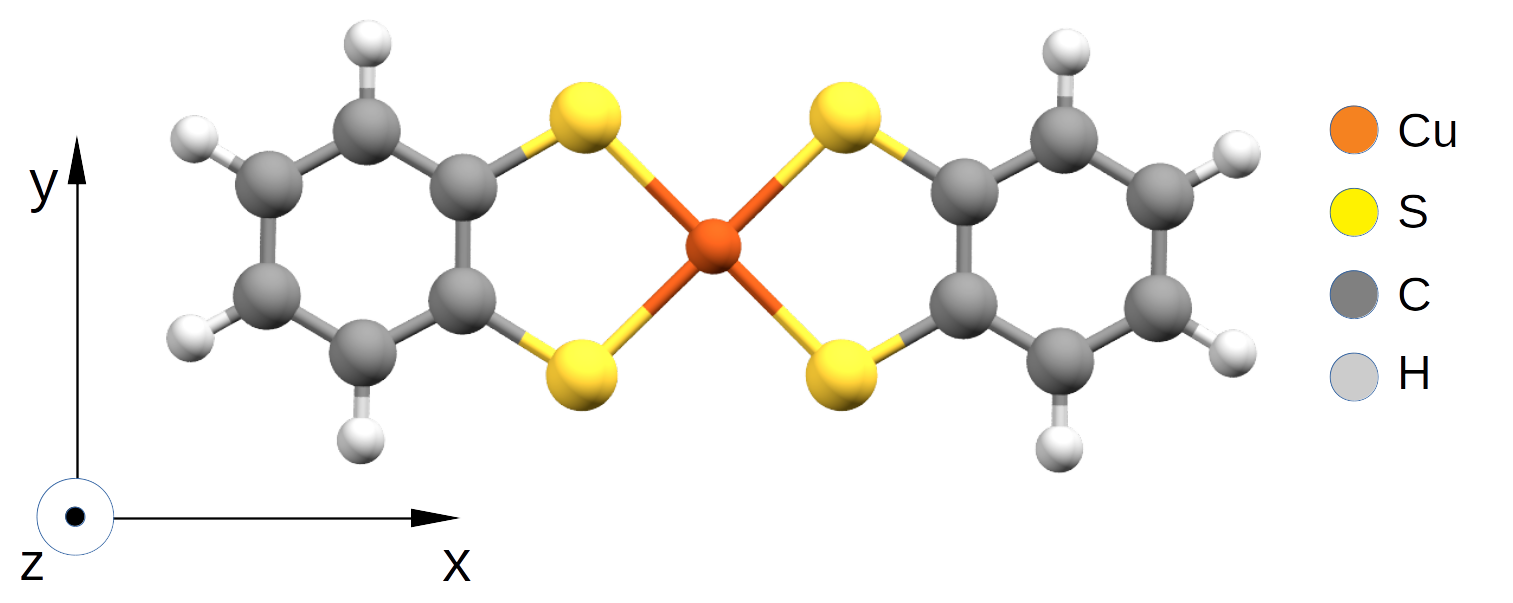}
    \caption{Molecular geometry of $\mathrm{[Cu(C_6 H_4 S_2)_2]^{2- }}$ (\textbf{Q-2}) from the experimental data \cite{Maiti2014}.
    The molecule is almost planar in the $xy$ plane. \textbf{Q-1} has a similar molecular geometry to \textbf{Q-2}.}
    \label{fig:Q2-geo}
\end{figure}

Figure~\ref{fig:Q2-geo} shows the molecular geometry of $\mathrm{[Cu(C_6 H_4 S_2)_2]^{2-}}$ (referred to as \textbf{Q-2}) derived from the
experimental data \cite{Maiti2014}. The molecule has almost planar structure in the $xy$ plane with inversion symmetry up to precision of
$\sim 10^{-4}$\AA.~There is an approximate $D_{2h}$ symmetry. The Cu-S bond lengths are 2.294 and 2.265~\AA, and the S-Cu-S angles are
89.6 and 90.4$^{\circ}$. A nominal ionic picture dictates that the Cu$^{2+}$ ion (3$d^9$) carries the spin $S=1/2$, while the four S atoms, each
of which has an oxidation state of $-1$, do not carry spin. However, the experimental data \cite{Fataftah2019,Maiti2014} strongly suggests covalent bonding between the Cu and the S atoms. In our study, we use the experimental geometry of \textbf{Q-2} \cite{Maiti2014} with C-H bond lengths
modified to be a standard value of 1.09~\AA, without further geometry relaxation. The coordinates of the molecular geometry are listed in Table~\ref{tab:Q2-geo-app} in the Appendix \ref{app:geo}.

The experimental geometry of the monoanionic molecule, $\mathrm{[Cu(C_6 H_4 S_2)_2]^{1- }}$ (referred to as \textbf{Q-1}), is not available. Therefore, we separately optimize the molecular geometry of \textbf{Q-1} for the triplet ($S=1$) and singlet state ($S=0$) without symmetry constraints until the root mean square of the force is less than 1~mHa/$a_B$ (where $a_B$ is Bohr radius) within the Perdew-Burke-Ernzerhof (PBE) generalized-gradient approximation (GGA) \cite{Perdew1996} using the NRLMOL code \cite{NRLMOL_1,NRLMOL_2,NRLMOL_3}. Our DFT calculation without SIC shows that the singlet state has a lower energy than the triplet state (by 0.39~eV). This trend agrees with the literature \cite{Maiti2014}. The optimized singlet structure has slightly reduced Cu-S bond lengths such as 2.223 and 2.228~\AA (compared to those for \textbf{Q-2}) with the S-Cu-S angles of 89.3 and 90.7$^{\circ}$. An experimental crystallographic data on a similar monoanionic Cu-based molecule, $\mathrm{[Cu(C_{14} H_{20} S_2)_2]^{1- }}$ suggests that the Cu-S bond lengths are 2.16 and 2.17~\AA~in the singlet state~\cite{Ray2005}, which agrees with the reported experimental Cu-S bond lengths in \textbf{Q-1} \cite{Maiti2014}. In our study, we use the PBE-GGA-optimized geometry
for \textbf{Q-1} that has inversion symmetry up to the precision of $\sim 10^{-4}$\AA~and approximate $D_{2h}$ symmetry. The coordinates of the optimized geometry are listed in Table~\ref{tab:Q1-geo-app} in the Appendix \ref{app:geo}.


\section{FLO-SIC Methods} \label{sec:methods}



\subsection{\label{theory} FLO-SIC formulation}

In order to include the SIC in an approximate exchange-correlation energy $E^{\rm{app}}_{\rm{xc}}[n_{\uparrow},n_{\downarrow}]$, Perdew and Zunger \cite{Perdew1981} considered the following self-interaction corrected exchange-correlation energy $E_{\rm{xc}}^{\rm{SIC}}$:
\begin{eqnarray}
    E_{\rm{xc}}^{\rm{SIC}} &=& E^{\rm{app}}_{\rm{xc}}[n_{\uparrow},n_{\downarrow}]
    - \sum_{i, \sigma} ( U^{\rm{s}}[n_{i \sigma}]+E^{\rm{app}}_{\rm{xc}} [n_{i \sigma}, 0] ), \label{eq:PZ_SIC} \\
    n_{\sigma}(\bm{r})&=& \sum_i |\phi_{i \sigma}(\bm{r})|^2 = \sum_{\alpha} |\psi_{\alpha \sigma}(\bm{r})|^2,
\end{eqnarray}
where $\phi_{i \sigma}(\bm{r})$ and $\psi_{\alpha \sigma}(\bm{r})$ are localized occupied orbitals and canonical (Kohn-Sham) orbitals, respectively,
and $n_{i \sigma={\uparrow,\downarrow}}(\bm{r})=|\phi_{i \sigma}(\bm{r})|^2$. By definition, the self-direct Coulomb energy $U^{\rm{s}}[n_{i \sigma}]$ of a single fully occupied electron must completely cancel its exact exchange-correlation energy $E_{\rm{xc}} [n_{i \sigma}, 0]$. However, the exact cancellation is not achieved within approximate exchange-correlation functionals, and Eq.~(\ref{eq:PZ_SIC}) takes care of the incomplete cancellation. The SIC energy $E_{\rm{xc}}^{\rm{SIC}}$ depends on both the electron density and the orbitals rather than just the electron density.

In the FLO-SIC method \cite{Pederson2014,Pederson2015,Yang2017}, in order to maintain unitary invariance of the SIC energy with respect to orbital transformations, Eq.~(\ref{eq:PZ_SIC}), a set of Fermi orbitals $F_{i\sigma}(\bm{r})$ are constructed from an initial set of Kohn-Sham orbitals $\psi_{\alpha\sigma}(\bm{r})$ as follows:
\begin{equation} \label{eq:FOs}
F_{i\sigma}(\bm{r}) = \frac{\sum_{\alpha}\psi_{\alpha\sigma} (\bm{a}_{i\sigma}) \psi_{\alpha\sigma}(\bm{r}) }{\sqrt{\sum_{\alpha}|\psi_{\alpha\sigma}(\bm{a}_{i\sigma})|^2}},
\end{equation}
where the index $i$ runs over all occupied orbitals. Here $\bm{a}_{i\sigma}$ are three-dimensional spatial coordinates assigned to each occupied orbital $i$ with spin $\sigma$, which are referred to as Fermi-orbital descriptor (FOD). Unitary invariance is assured since any set of orthonormal orbitals that span the occupied space can be used in Eq.~(\ref{eq:FOs}). Note that chemical properties of systems are reflected in the optimized FODs. Equation~(\ref{eq:FOs}) ensures localization and the L\"owdin scheme \cite{Lowdin1950} guarantees orthonormality of orbitals. The constructed orbitals are referred to as Fermi-L\"owdin orbitals (FLO). The localized orbitals are used to ensure size consistency of the SIC energy.
The total energy including the SIC energy must be minimized both with respect to the density, through standard self-consistent calculations, and with respect to FOD positions.



The FLO-SIC calculations are performed using the FLOSIC 0.2 program \cite{FLOSIC_code}, which is based on the NRLMOL code \cite{NRLMOL_1,NRLMOL_2,NRLMOL_3}. We use the Gaussian-type basis sets optimized by Pederson and Porezag \cite{Porezag1999}. Typically,
a SIC-free DFT calculation within the L(S)DA-PW92 \cite{Perdew1992} or PBE-GGA exchange-correlation functional is first carried out to obtain the converged electron density and an initial set of Kohn-Sham orbitals. Then for the given electron density, an initial set of FLOs is constructed
using the Kohn-Sham orbitals and an initial set of FODs [Eq.~(\ref{eq:FOs})]. Next the SIC energy is computed and the electron density is updated
in a self-consistent-field (SCF) loop within the L(S)DA-PW92 exchange-correlation functional including the one-electron SIC potentials. Using the updated electron density, the set of the FOD positions is updated with a gradient-based optimization algorithm using analytic FOD forces \cite{Pederson2015}. This procedure is repeated until the total energy including the SIC energy reaches a minimum and the maximum force on the FODs is less than a tolerance.

\subsection{\label{sec:frozen}Frozen-density loop method}



\begin{figure}[htb!]
    \centering
    \includegraphics[width=0.6\linewidth]{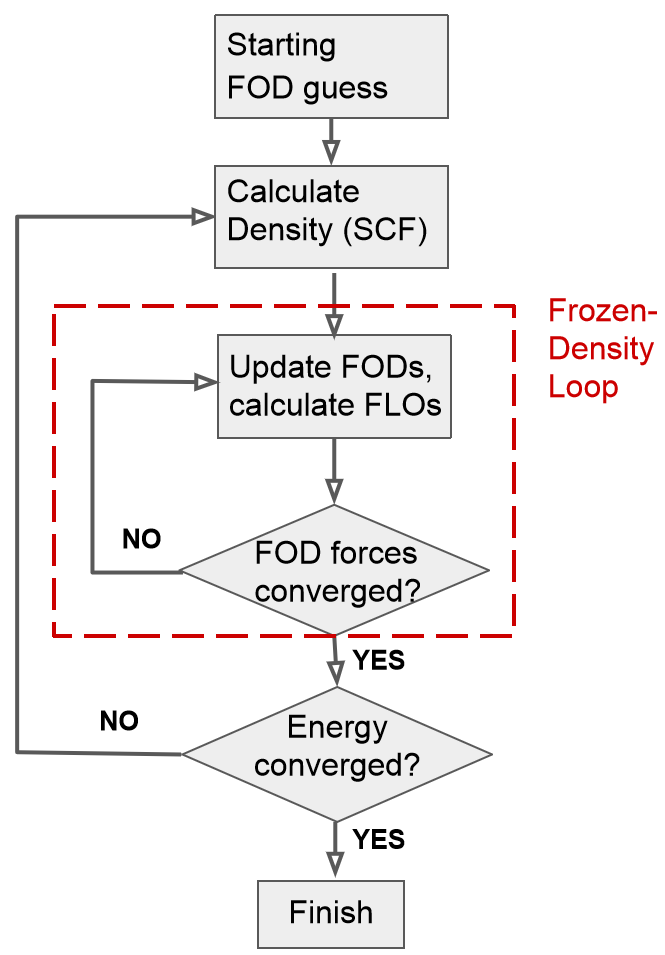}
    \caption{Schematic flow chart of the frozen-density loop procedure within the FLO-SIC method.}
    \label{fig:frozen}
\end{figure}

For large molecules especially including transition-metal elements, many starting sets of FODs may need to be considered in FLO-SIC calculations, and the convergence of the total energy with respect to the FODs is known to be much slower than the electronic convergence of a SIC-free DFT SCF run. Therefore, there is a great demand for expediting FLO-SIC calculations. We apply the following modification to the FLO-SIC method. Instead of re-optimizing the electron density within a SCF loop after every FOD update, the density is held fixed (frozen) while the SIC energy is minimized with respect to FOD positions only. When the FOD geometry converges with the given FOD force tolerance (0.5~mHa/$a_B$) in our calculations, the electron density is recalculated self-consistently with the new converged FOD positions, and the procedure is repeated until the total energy converges. 
In this method referred to as frozen-density loop algorithm (Fig.~\ref{fig:frozen}), a full FOD optimization is performed for each SCF density. 
We find the frozen density scheme to be significantly more efficient, saving up to a factor of six in total computational effort for the dianonic Cu-based molecule consisting of 175 electrons. The saving does not result from a decrease in the number of steps in the gradient optimization of 
the FODs, which is roughly the same in each approach. Instead, it derives from a much smaller number of SCF steps taken in the frozen density approach.

\subsection{\label{sec:initial-Q2} Starting sets of FODs for the Q-2 molecule}

Multiple local minima in total energy may be possible in the FOD optimization \cite{DY_Kao}. In order to ensure that we reach the minimum-energy FOD configuration, we consider multiple starting sets of FODs described in this subsection. Since the number of three-dimensional FOD position vectors equals the number of electrons, we need 175 initial FOD vectors for the \textbf{Q-2} molecule which consists of 88 spin-up (majority-spin) and 87 spin-down (minority-spin) FODs. Previous FLO-SIC studies showed that starting sets of FODs of ligands (i.e., $s$- and $p$-electron systems) generated by the Monte-Carlo-based fodMC code \cite{Schwalbe2018} converge rapidly to the minimum SIC energy, whereas effectiveness of the fodMC code for transition-metal systems is not guaranteed. FOD positions associated with or around the transition-metal centers are also known to converge 
extremely slowly despite the fact that these FOD forces are largest. Therefore, for the starting sets of FODs, it makes sense to treat Cu and 
ligand FODs separately.


We assign the initial FODs of the ligands using the fodMC code. Considering the Cu-S bonding, we place four spin-up and four spin-down FODs at the same positions along the Cu-S bonding directions. This assignment is consistent with the majority of the spin density from the Cu center. There is one ambiguity related to the delocalized $p$-electrons on the benzene-like rings. One obvious way to put FODs there is to follow a classical single-double bond picture, and assign one FOD of each species to each single bond and two FODs to each double bond. Another way is to alternate, such that one C-C bond in the ring has 2 spin-up FODs and 1 spin-down FOD, and the next bond has 1 spin-up FOD and 2 spin-down FOD. The latter alternating picture is related to Linnett double-quartet theory and has been shown to achieve slightly lower energy than the former scheme \cite{FLOSIC_Linnett}. Therefore, the alternating FOD pattern is used for the ligands. These starting ligand FODs are used for all the FOD configurations of the same charge state, which differ only in the Cu starting FODs.

Since the fodMC program does not guarantee good starting FOD positions for transition-metal centers, a different strategy should be used.
One possibility is to use a set of pre-converged single-atom results by Kao et.al \cite{DY_Kao}. Although this could be a good starting point, it has several disadvantages. Firstly, the electronic structure of the bonded, oxidized atom differs from that of an isolated one, and so significant re-optimization of the starting positions may be expected. Secondly, it is not clear which FODs from the optimized neutral atom should be removed 
to create the expected oxidation state. Finally, this approach provides few possibilities to systematically probe the space of initial FOD geometries.

Another, more systematic approach is based on the following considerations. It is known from single-atom results \cite{DY_Kao} that FODs tend to group by the principal quantum numbers of orbitals they represent. For instance, in Ar atom, for each spin channel there is one FOD at the center corresponding to 1$s$ orbital, 4 FODs in a shape of tetrahedron centered at the atom (for 2$s$ and 2$p$ orbitals), and another 4 FODs in a tetrahedron but of larger dimensions for 3$s$ and 3$p$ orbitals. The molecular symmetry needs to be reflected in the initial FOD geometries.
Considering that the Cu-S bond angles are close to $90^{\circ}$, there is an approximate 4-fold rotational symmetry of Cu atom with respect to Cu-S bonds.

\begin{figure}[h!]
    \centering
    \includegraphics[width=0.85\linewidth]{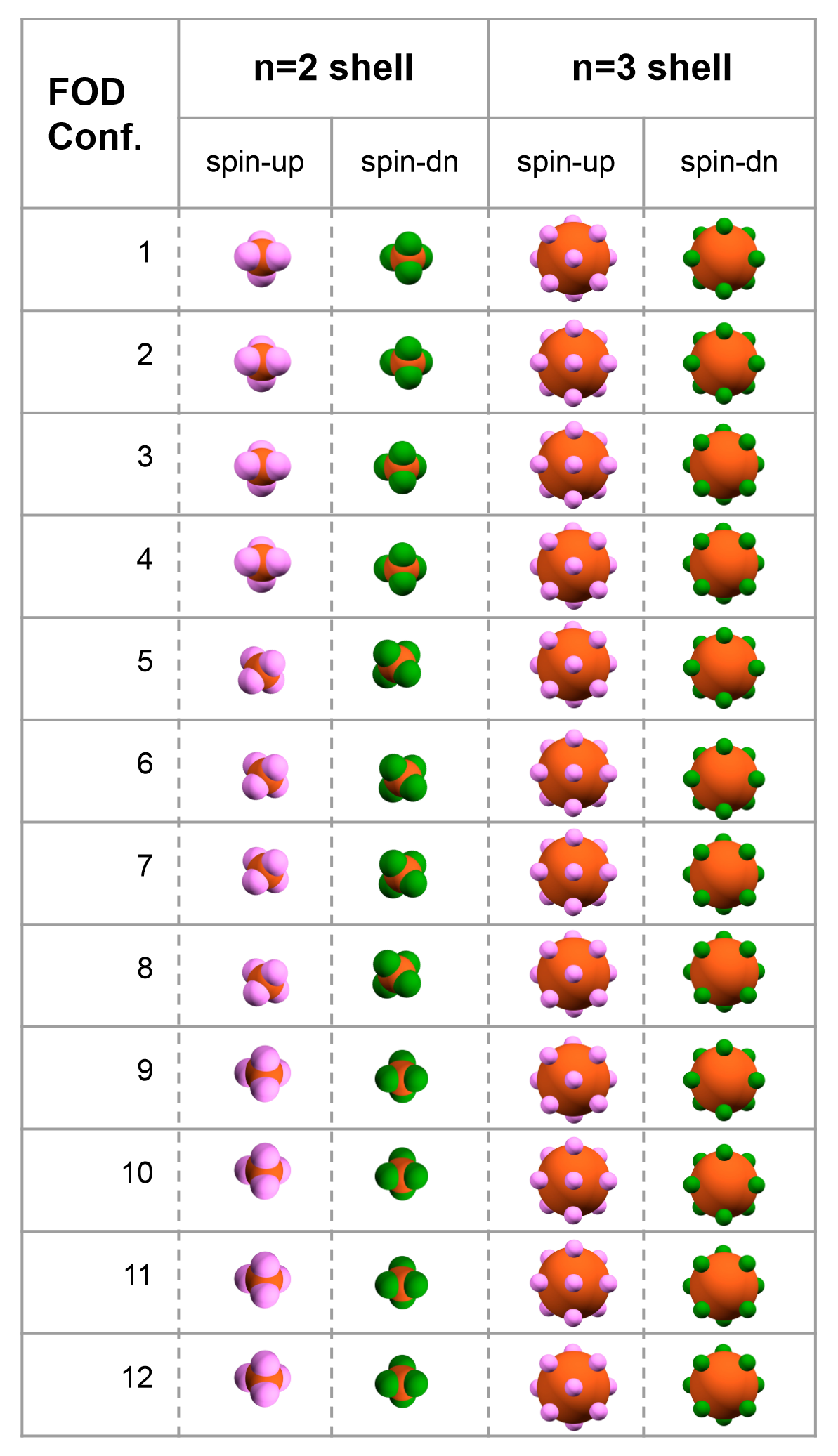}
    \caption{Schematic depiction of 12 starting FOD geometries associated with Cu in the \textbf{Q-2} molecule. Orange spheres represent concentric spheres around the Cu atom where magenta (green) dots are for spin-up (spin-down) FODs.}
    \label{fig:flgeoms}
\end{figure}

Based on the above considerations, the following procedure is used for systematic generation of 12 initial Cu FODs (see Fig.~\ref{fig:flgeoms}):
\begin{itemize}
    \item Identify three groups of FODs around the Cu center, distinct by the their distances to the center, following the single-atom Cu FOD result \cite{DY_Kao}.
    \item Using these distances, consider the following number of FODs at concentric spherical shells: 1 spin-up and 1 spin-down FOD at the center ($n=1$); 4 spin-up and 4 spin-down FODs at a smaller ($n=2$) sphere; and 9 spin-up and 8 spin-down at the larger ($n=3$) sphere.
    \item Place these FODs in a 4-fold symmetric fashion as shown in Fig. \ref{fig:flgeoms}.
    For the $n=2$ shell, 4 FODs are placed at the vertices of opposing tetrahedra, in each of which two edges are parallel to the molecular plane either oriented along the $x$ and $y$ axes or along the diagonals (i.e. Cu-S bonds);
    For the $n=3$ shell, 4 FODs are above the molecular plane, forming a square, while 4 FODs are below the plane in a rhombus, with 1 spin-up FOD directly on top of the Cu site.
\end{itemize}

\subsection{\label{sec:initial-Q1} Starting sets of FODs for the Q-1 molecule}

\begin{figure}[h!]
    \centering
    \includegraphics[width=0.85\linewidth]{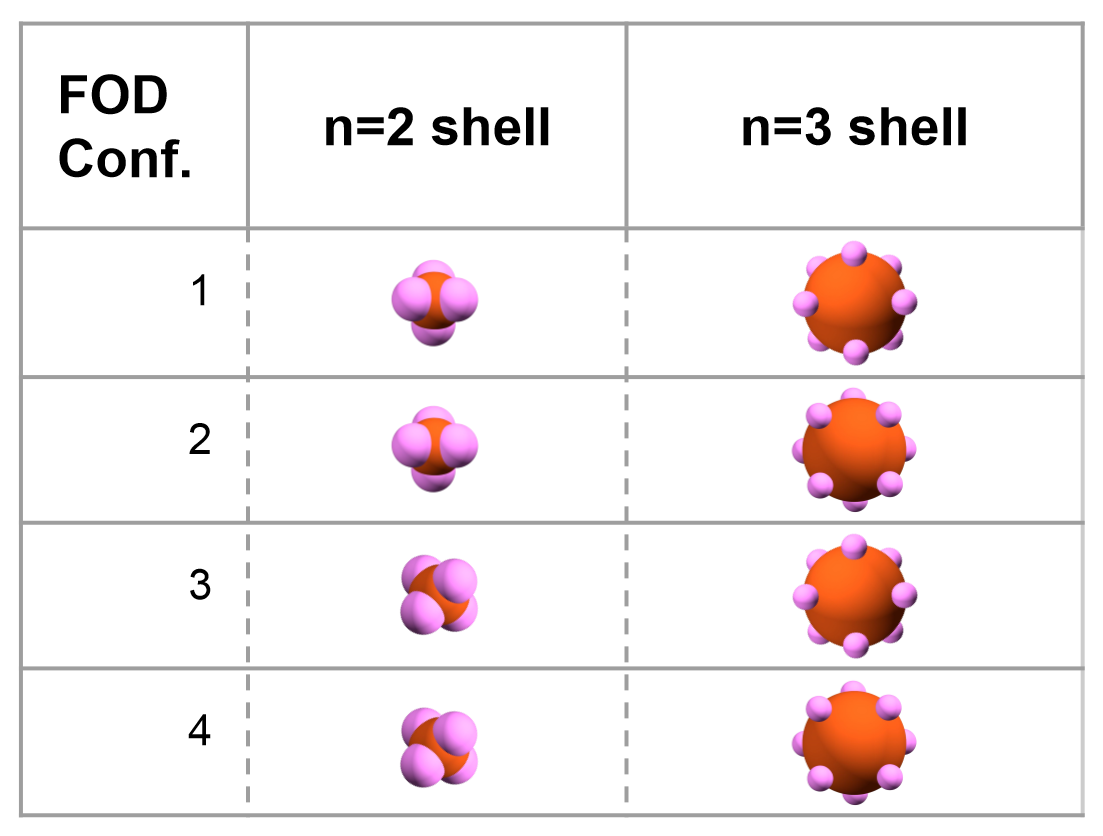}
    \caption{
    Schematic depiction of 4 starting spin-unpolarized FOD geometries associated with Cu in the \textbf{Q-1} molecule. Orange spheres represent concentric spheres around the Cu atom where magenta dots are for FODs.}
    \label{fig:FOD-Q1}
\end{figure}

For generation of initial FODs for the \textbf{Q-1} molecule, a similar scheme to that of the \textbf{Q-2} molecule is used. The initial Cu FODs
that we consider are shown in Fig.~\ref{fig:FOD-Q1}. Since the \textbf{Q-1} molecule is in a singlet state, we perform spin-unpolarized FLO-SIC calculations, where spin-up and spin-down FODs must occupy the same positions. This reduces the number of FODs and computational time by half. The results shown in Sec.~\ref{sec:results-Q1} are obtained using spin-unpolarized FLO-SIC calculations. In all other aspects, the procedure on the \textbf{Q-1} is the same as on the \textbf{Q-2} molecule.


\section{\label{sec:results-Q2}FLOSIC Results for the \textbf{Q-2} Molecule}

\subsection{Converged FODs and energy convergence}

\begin{figure}[h!]
\centering
\includegraphics[width=0.7\linewidth]{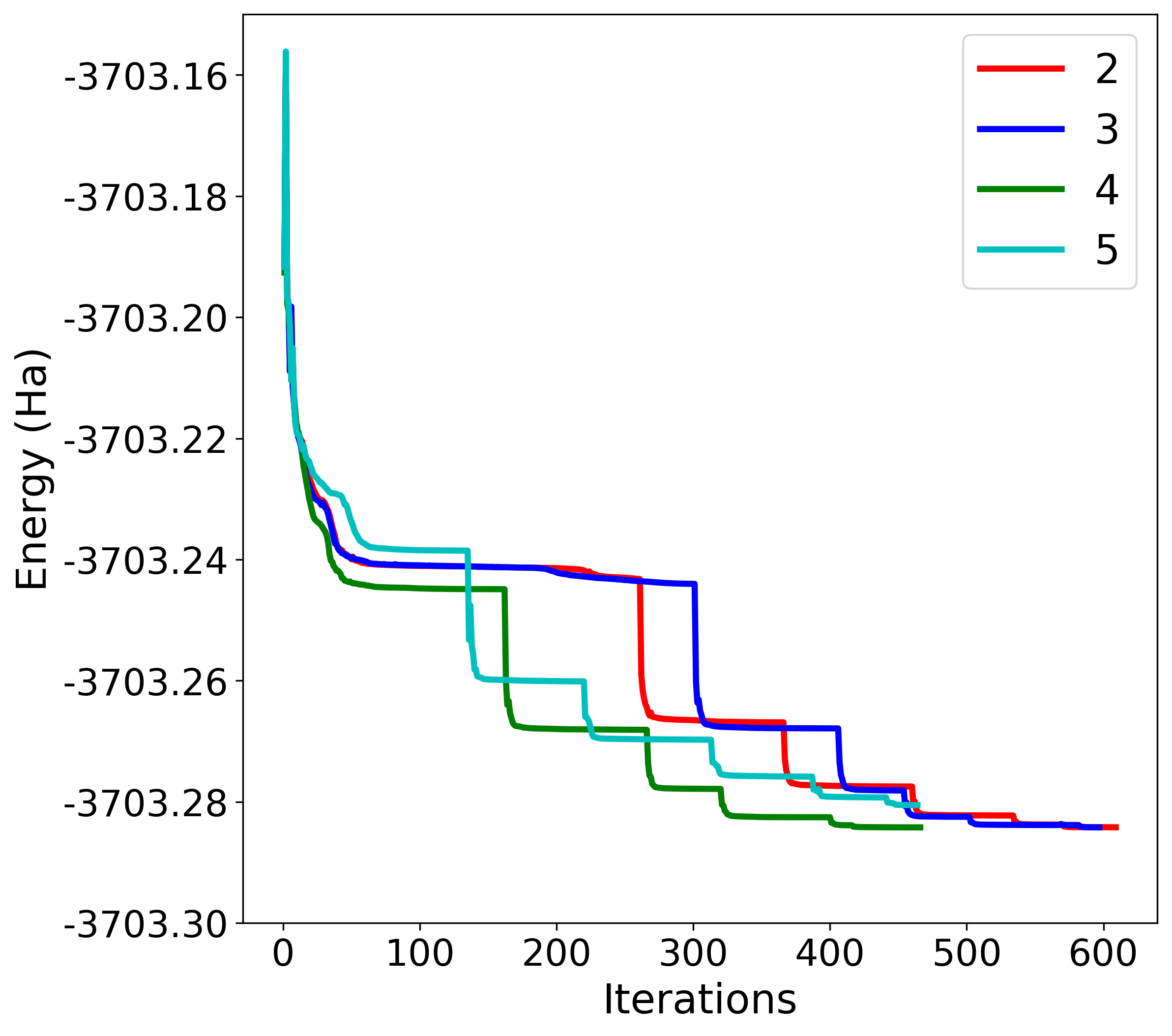}
\caption{Total energy versus FOD update or iteration number for FOD configuration 2-5 (labeled in Fig.~\ref{fig:flgeoms}) of the \textbf{Q-2} molecule. After the initial steep relaxation of the energy, the step-function-like abrupt jump occurs whenever the electron density is updated
after each frozen-density loop converges. Here 6 frozen-density loop cycles are shown.}
\label{fig:Q-2energ}
\end{figure}

For the 12 starting FOD configurations (Fig.~\ref{fig:flgeoms}) with the same DFT-converged electron density, FLO-SIC calculations are carried out using the frozen-density loop method (Fig.~\ref{fig:frozen}). After the first frozen-density loop cycle, converged FODs starting from configuration 9-12 are found to have similar total energies and similar FOD positions to FOD configuration 1-4, respectively. After the second frozen-density loop cycle, FOD configuration 1 is found to reach the same converged energy and same FOD positions as FOD configuration 4. Therefore, we continue to relax the FOD positions and the electron density only for the remaining 7 FOD configurations (configuration 2-8). After completion of 6 cycles of the frozen-density loop, we find that the SCF total energy converges within below or up to 1 mHa. Figure~\ref{fig:Q-2energ} shows the convergence of the SCF total energy versus FOD iteration number for configuration 2-5. The sudden drops in the energy correspond to switches between frozen-density loop cycles. Although different FOD configurations converge at different rates, configuration 2-4 seem to converge to the same total energy.

\begin{table}[h!]
\caption{Converged SCF total energies $E$ and maximum final force components $F_{\rm{max}}$ after the 6 frozen-density loops, and energy differences $\Delta E$ between the 6-th and 5-th frozen-density loop cycles for the configuration 2-8 (labeled in Fig.~\ref{fig:flgeoms}) of the \textbf{Q-2} molecule. Each frozen-density loop cycle consists of many FOD updates. For example, for the FOD configuration 2, there are 609 updates or iterations of the FODs after the 6 frozen-density loop cycles.}
\begin{ruledtabular}
\begin{tabular}{c c c c}
FOD  & $E$ (Ha) & $F_{\rm{max}}$   & $\Delta E$  \\
Conf &          & (mHa/$a_{\rm B}$) & (mHa)      \\ \hline
2 & -3703.28420 & 0.48 & -0.43  \\
3 & -3703.28421 & 0.48 & -0.37  \\
4 & -3703.28422 & 0.36 & -0.35  \\
5 & -3703.28054 & 0.37 & -1.30  \\
6 & -3703.28189 & 0.32 & -0.53  \\
7 & -3703.28401 & 0.42 & -0.77  \\
8 & -3703.28273 & 0.39 & -0.60
\end{tabular}
\end{ruledtabular}
\label{tab:Q2-energy}
\end{table}


The total energy changes by about 1~mHa or less than 1~mHa after the 6-th frozen-density loop cycle compared to the energy after the 5-th cycle
(see the $\Delta E$ values in Table~\ref{tab:Q2-energy}). Therefore, we analyze our FLO-SIC results using the data obtained after the 6-th cycle. Each cycle consists of many FOD updates or iterations as shown in Fig.~\ref{fig:Q-2energ}. Table~\ref{tab:Q2-energy} lists the converged total energies and maximum force components after the 6 frozen-density loop cycles. FOD configuration 4 gives the lowest energy, but the energy differences among the different FOD configurations are on the order of mHa at most. The maximum force component $F_{\rm{max}}$ is about 0.3-0.5 mHa/$a_{\rm B}$, while the maximum electric dipole moment component is about 0.006-0.007 in atomic units (and the maximum dipole moment component from the SIC-free PBE-GGA is
about 0.001 in atomic units).

\begin{figure}[h!]
\centering
\includegraphics[width=0.9\linewidth]{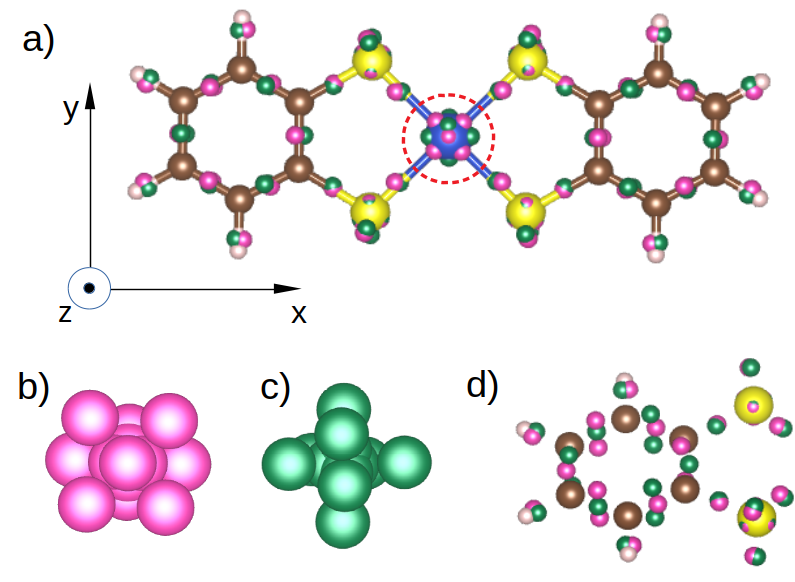}
\caption{(a) Final converged FOD positions for configuration 4 of the \textbf{Q-2} molecule with magenta (green) color for spin-up (spin-down) FODs.
(b) and (c) Zoom-in of the spin-up and spin-down FODs near the Cu atom. (d) Zoom-in of the spin-up and spin-down FODs near the ligands. The zoom-in region is indicated as a red dashed circle in (a).}
\label{fig:Q-2_Final_FODs}
\end{figure}

Figure~\ref{fig:Q-2_Final_FODs} shows the converged FOD positions of configuration 4 after the 6-th frozen-density loop cycle. The initial spin-down FOD positions significantly change upon relaxation compared to the initial spin-up FOD positions. The converged spin-up and spin-down FOD positions (starting from configuration 4) are listed in Tables~\ref{tab:Q2-FOD-up} and \ref{tab:Q2-FOD-dn} in the Appendix \ref{app:Q2-FODs}. Henceforth, we examine the FLO-SIC calculated electronic structure of the molecule using the electron density obtained from this converged FOD configuration.
We confirm that the FLO-SIC electronic structure from the other converged FOD configurations 2, 3, 5, 6, 7, and 8 is very close to that from the converged FOD configuration 4, within our numerical accuracy (Table~\ref{tab:Q2-dep-HL} in the Appendix), although the final configurations of the 
core and valence FODs somewhat differ from one another. We also check that the electronic structure does not change with further relaxations of the FODs beyond the threshold and/or with further decrease of the electric dipole moment.

\subsection{FLOSIC-calculated electronic structure}


Using the FLO-SIC result, we calculate Mulliken spin populations of the Cu atom, all four S atoms, and all C atoms of the \textbf{Q-2} molecule ($S=1/2$). We find that the majority of the spin density (67\%) arises from the Cu center and the rest from the S atoms (33\%) and from the C atoms ($-$3\%). These FLO-SIC calculated values are close to the values from our UHF calculation (Table~\ref{tab:SpinPop}). Although electron correlation is not included in the UHF calculation, there is no SIE in it. Therefore, this indicates that the SIC is properly taken care of in our FLO-SIC calculation. The FLO-SIC spin populations are also close to our multiconfigurational complete active space self-consistent field (CASSCF) result,
as well as a previous multireference result \cite{Fataftah2019} based on complete active space second-order perturbation theory (CASPT2). Since
CASPT2 includes both static and dynamic correlations, our FLO-SIC result is very encouraging. The detail of our CASSCF and UHF calculations is described in the Appendix \ref{app:caspt2}.

Now we compare the FLO-SIC spin populations to those from our SIC-free DFT calculations (Table~\ref{tab:SpinPop}). Both LSDA-PW92 and PBE-GGA exchange correlation functionals without the SIC give rise to much more delocalized spin density with a large contribution from the S atoms (about 70\%) and a much smaller contribution from the Cu center (about 30\%).

\begin{table}[h!]
\caption{Calculated Mulliken spin populations of the Cu and all four S atoms (in units of Bohr magneton $\mu_{\rm B}$) of the \textbf{Q-2} molecule ($S=1/2$) using different levels of computational methods. In the CASPT2 calculation \cite{Fataftah2019}, the C spin population was not reported, 
while in the EPR experiment \cite{Maiti2014}, only the spin density from the Cu $d_{xy}$ orbital was reported. The Cu, S, and C spin populations 
from the FLO-SIC and SIC-free DFT calculations do not add up to 1.00~$\mu_{\rm B}$ due to very small spin populations on the H atoms. All 
of the values are our results unless specified otherwise.}
\label{tab:SpinPop}
\begin{ruledtabular}
\begin{tabular}{lccc}
Method &  Cu  &  S  &  C \\ \hline
FLO-SIC                      &  0.67  &  0.33  & -0.03 \\
Unrestricted Hartree Fock    &  0.79  &  0.22  & -0.01 \\
CASSCF(11,11)                &  0.70  &  0.29  & 0.01 \\
CASPT2 (Ref.\onlinecite{Fataftah2019})&  0.76  &  0.24 & N/A \\
SIC-free DFT (LSDA-PW92)     &  0.31  &  0.72  & -0.08 \\
SIC-free DFT (PBE-GGA)       &  0.32  &  0.73  & -0.12  \\
B3LYP DFT (Ref.\onlinecite{Maiti2014}) & 0.24  & 0.76  & 0.00 \\
EPR experiment (Ref.\onlinecite{Maiti2014}) & 0.51 & N/A & N/A
\end{tabular}
\end{ruledtabular}
\end{table}

We compute the energy levels of the HOMO and LUMO using the FLO-SIC result (see Table~\ref{tab:homo}) for the \textbf{Q-2} molecule, finding that the
HOMO energy is negative. This indicates that the dianionic form of the molecule can exist, which is consistent with the experimental synthesis of the
dianionic \textbf{Q-2} molecule \cite{Fataftah2019,Maiti2014}. The SIC-free PBE-GGA calculation provides a positive HOMO energy. The comparison between the FLO-SIC and the SIC-free DFT results shows that the SIC shifts the HOMO level downward by 5.07 eV, while it shifts the LUMO level upward by 0.71 eV. The FLO-SIC HOMO-LUMO gap is about 6.34~eV, while the corresponding gap from the SIC-free DFT is about 0.57~eV. Since the SIC potential energy is typically negative, it lowers the energies of the occupied orbitals. As a result, the HOMO-LUMO gap increases with the SIC. The small change of the LUMO level with the SIC is due to the orbital relaxation effect.


\begin{table}[h!]
\caption{Contributions of the Cu $d$, all S $p$, and all C $p$ orbitals to the HOMO and LUMO of the \textbf{Q-2} (charge $Q=-2$, $S=1/2$)
molecule calculated using the FLO-SIC method in comparison to our SIC-free PBE-GGA calculations as well as an earlier B3LYP result
\cite{Maiti2014} where the orbital decomposition was not reported. In our calculations (PBE-GGA, FLO-SIC), the HOMO arises from the
spin-up (majority-spin) orbital and the LUMO from the spin-down (minority-spin) orbital (see Fig.~\ref{fig:Q2-orbs}), while in the
B3LYP calculation \cite{Maiti2014}, both HOMO and LUMO are from the spin-down orbitals.}
\label{tab:homo}
\begin{ruledtabular}
\begin{tabular}{cccrrr}
Method &  Level  &  Energy(eV) & Cu $d$ & S $p$ &  C $p$ \\ \hline
FLO-SIC & HOMO & -1.99 & 2.2\% &74\% &23\% \\
        & LUMO & 4.35 & 85\% &11\% &0.5\% \\ \hline
PBE-GGA & HOMO & 3.08 & 49\% &46\% &1.4\% \\
        & LUMO & 3.65 & 57\% &37\% &1.5\% \\ \hline
 B3LYP \cite{Maiti2014} & HOMO & $\sim$3.3 & - & - & - \\
        & LUMO & $\sim$5.3 & - & - & -
\end{tabular}
\end{ruledtabular}
\end{table}

\begin{figure*}[th!]
\centering
\includegraphics[width=0.9\textwidth]{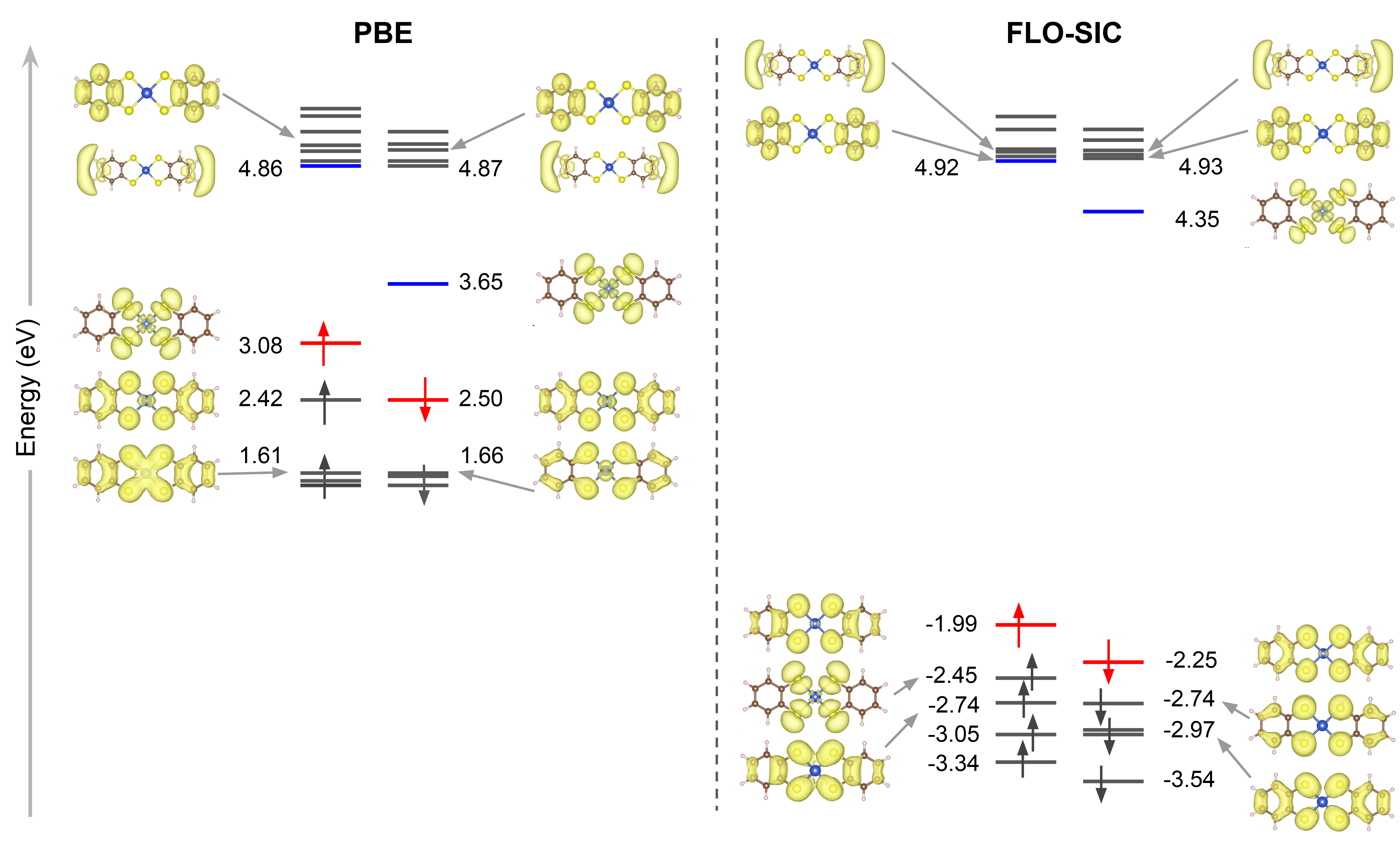}
\caption{Energy levels and corresponding canonical orbitals of the \textbf{Q-2} molecule from the SIC-free PBE-GGA and FLO-SIC calculations}
\label{fig:Q2-orbs}
\end{figure*}

\begin{figure*}[th!]
\centering
\includegraphics[width=0.75\linewidth]{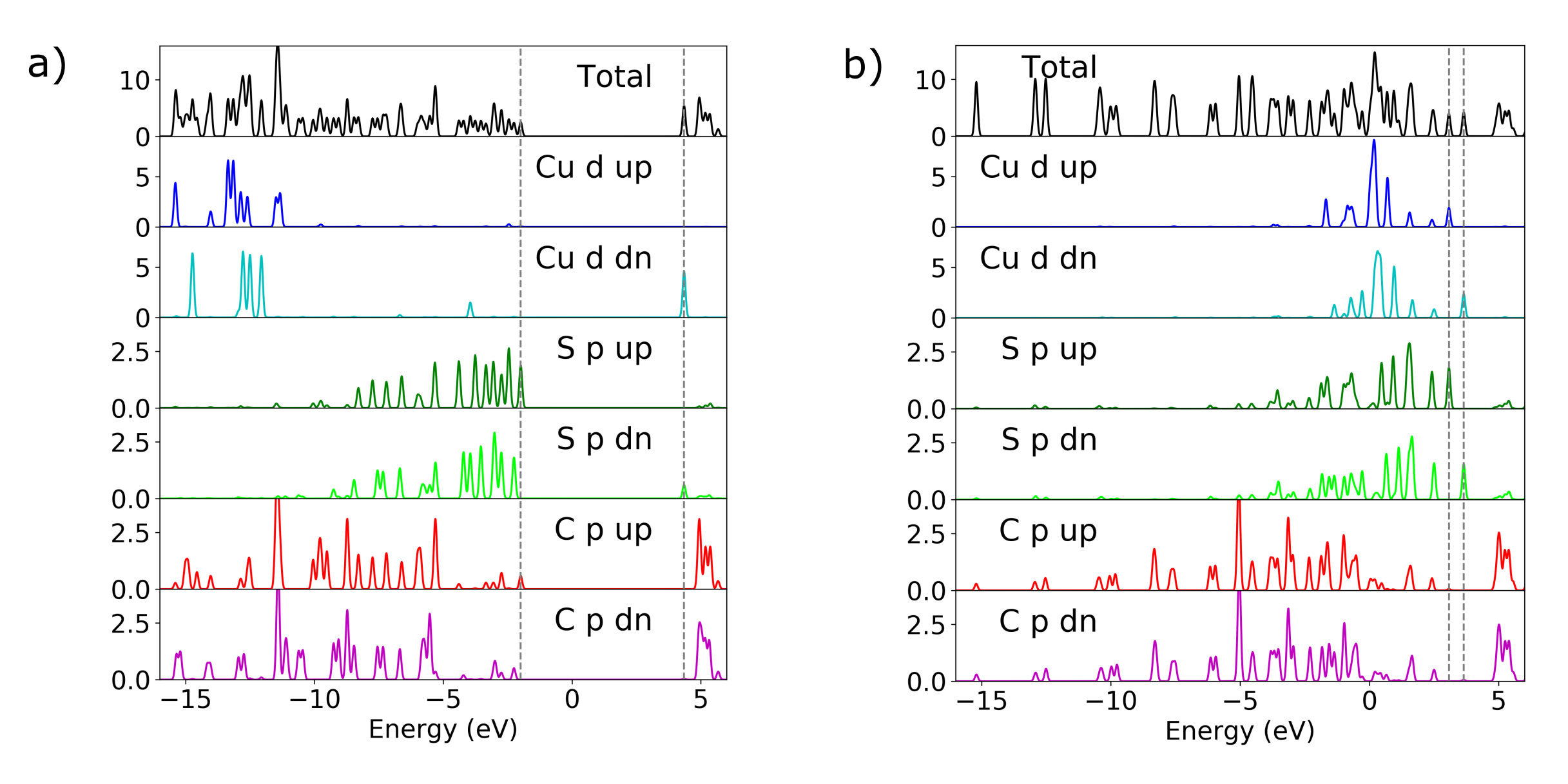}
\caption{(a) FLO-SIC and (b) SIC-free DFT calculated densities of states (DOS) projected onto spin-up and spin-down Cu $d$, S $p$, and C $p$ orbitals of the \textbf{Q-2} molecule where the left and right vertical dashed lines indicate the HOMO and LUMO levels. The FLO-SIC HOMO-LUMO gap is 6.34 eV, while the SIC-free DFT HOMO-LUMO gap is 0.56 eV.}
\label{fig:Q2-dos}
\end{figure*}

Figure~\ref{fig:Q2-orbs} shows the HOMO and LUMO and a few other {\it canonical} orbitals of the \textbf{Q-2} molecule calculated from the FLO-SIC method in comparison to the PBE-GGA orbitals. The FLO-SIC HOMO looks similar to the SIC-free DFT HOMO-2, while the FLO-SIC HOMO-2 looks similar to 
the SIC-free DFT HOMO. The other occupied orbitals do not seem to change their overall shapes other than the energies and Cu contributions. Table~\ref{tab:homo} lists quantitative characteristics of the HOMO and LUMO using the FLO-SIC method and the PBE-GGA. The FLO-SIC
calculation shows that the HOMO arises mainly from the spin-up S $p_{x,y}$ orbitals with significant contributions from the spin-up C $p_{x,y}$ orbitals and a tiny contribution from the spin-up Cu $d_{xy}$ orbitals, whereas the LUMO consists of a major contribution from the spin-down Cu $d_{xy}$ and some contributions from the spin-down S $p_{x,y}$ orbitals. These orbital characteristics are very different from those obtained using 
the SIC-free DFT. For the latter, both the HOMO and LUMO consist of large contributions from the Cu $d_{xy}$ and S $p_{x,y}$ orbitals. The FLO-SIC 
LUMO and the SIC-free DFT HOMO and LUMO shown in Fig.~\ref{fig:Q2-orbs} can be identified as antibonding orbitals which are combinations of the Cu $d_{xy}$ and the S $p_{x,y}$ orbitals. The CASSCF calculation shows that the singly occupied active orbital of the lowest-energy configuration 
has the character of the majority-spin (spin-up) Cu-S antibonding orbital (molecular orbital 6 in Fig.~\ref{fig:casorbs}) and it can be viewed as the HOMO at the single-electron picture.

In order to understand the effect of FLO-SIC on the HOMO and other occupied orbitals, we plot the densities of states (DOS) projected onto spin-up and spin-down Cu $d$, S $p$, and C $p$ orbitals, as shown in Fig.~\ref{fig:Q2-dos}. From the comparison between the FLO-SIC and the SIC-free DFT DOS plots, we find that the SIC lowers the occupied S $p$ and C $p$ orbitals by about 5 eV and the occupied Cu spin-up 3$d$ orbitals by about 14-15~eV.
The SIC effect is expected to be stronger for the 3$d$ orbitals than for the $p$ orbitals since the 3$d$ orbitals are more strongly localized, i.e., larger self-interaction energy in the SIC-free DFT result. The highest occupied spin-down Cu $d$ orbital energy is lowered by only about 6.5 eV, 
which is much smaller than the case of the corresponding spin-up orbital energy due to hybridization of the spin-down Cu orbital with the spin-down S $p$ orbital. This is consistent with the PZ-SIC results of the 3$d$ transition metal atoms \cite{Perdew1981} which are attributed to a more attractive exchange-correlation potential seen by the spin-up 3$d$ orbitals. The characteristics of the HOMO qualitatively changes with the SIC because of the much larger downward shift of the spin-up Cu $d$ orbitals than the S $p$ orbitals.

\subsection{\label{sec:comp-Q2} Comparison with previous work}

Experimental data~\cite{Fataftah2019} on the \textbf{Q-2} molecule indicates a strong covalent nature of Cu-S bonding which leads to long spin relaxation and coherence times. EPR experiments~\cite{Maiti2014} imply that about 51\% of the spin density arises from the Cu $d_{xy}$ orbitals for \textbf{Q-2}. The CASPT2 calculation from Ref.~\onlinecite{Fataftah2019} shows that 76\% (24\%) of the spin density originates from the Cu (S) atoms. Both the experimental data and the CASPT2 result suggest a majority contribution of the Cu $d$ orbitals to the spin density, which is in line with our FLO-SIC and CASSCF spin populations. Especially, the CASPT2~\cite{Fataftah2019} and our CASSCF spin populations quantitatively agree with our FLO-SIC values (see Table~\ref{tab:SpinPop}).

However, recent DFT calculations \cite{Maiti2014} using a hybrid functional such as Becke, 3-parameter, Lee–Yang–Parr (B3LYP) \cite{Becke1993,Lee1988,Vosko1980,Stephens1994}, indicate that the Cu $d$ (S $p$) orbitals carry 24\% (76\%) of the total spin density and that both the HOMO and LUMO are from the spin-down (minority-spin) channel.The relatively smaller Cu contribution to the spin density (Table~\ref{tab:SpinPop}) and the positive HOMO energy (Table~\ref{tab:homo}) are consistent with our SIC-free PBE-GGA result. However, these B3LYP results mostly do not agree even qualitatively with the experimental data or our FLO-SIC results. Note that our CASSCF result is not in line with the B3LYP result, either. In order to reconcile the discrepancy between the EPR experimental data \cite{Maiti2014} and the B3LYP result and to explain the oxidation pathway to \textbf{Q-1}, Ref.~\onlinecite{Maiti2014} alternatively proposed co-existence of the two competing iso-electronic states for \text{Q-2} using TDDFT: $\mathrm{[Cu^{I}(bdt_{2},4S^{3-\star})]^{2-}}$ and $\mathrm{[Cu^{II}(bdt_{2},4S^{4-})]^2}$. In the former, the four S atoms give rise to spin $S=1/2$, while in the latter the Cu atom carries one unpaired electron. Our FLO-SIC results and multiconfigurational/multireference calculations as well as experimental data \cite{Fataftah2019} support the latter state rather than the proposed co-existent states.

\section{\label{sec:results-Q1}FLO-SIC Results for the \textbf{Q-1} Molecule}

Our SIC-free DFT calculations and experiments \cite{Maiti2014} on \textbf{Q-1} suggest that the singlet state has a lower energy than the triplet state. Therefore, we consider only the singlet state in the FLO-SIC calculations.

\subsection{Converged FODs}


The FLO-SIC procedure for studies of the \textbf{Q-1} molecule is similar to that described for \textbf{Q-2}. Since we carry out spin-unpolarized
FLO-SIC calculations, the total number of FODs is now reduced to 87. We start with 4 initial spin-unpolarized FOD configurations as illustrated in Fig.~\ref{fig:FOD-Q1}, similarly to the case of \textbf{Q-2}. We converge the 4 initial FOD configurations through 6 individual frozen-density loop cycles. Table~\ref{tab:Q1-energy} lists the converged total energies and maximum force components after the 6 cycles. FOD configuration 2 gives the lowest energy, but the energy differences among the different FOD configurations are on the order of 0.1~mHa. The maximum force component $F_{\rm{max}}$ is about 1~mHa/$a_{\rm B}$, while the maximum electric dipole moment component is about 0.009-0.01 in atomic units. The dipole moment is slightly larger than that for the \textbf{Q-2} molecule, which is consistent with the fact that the molecular geometry of \textbf{Q-1} was optimized without symmetry constraints. The converged FODs of configuration 2 are shown in Fig.~\ref{fig:Q1-FOD} and the FOD positions are listed in the Appendix \ref{app:Q1-FODs} (Table~\ref{tab:Q1-FODs-app}). We confirm that the FLO-SIC electronic structure from the other converged FOD configurations 1, 3, 
and 4 is very close to that from the converged FOD configuration 2, within our numerical accuracy (see Table~\ref{tab:Q1-dep-HL} in the Appendix \ref{app:FOD-comp}), although the final FOD configurations are somewhat different from one another. Henceforth, we analyze the FLO-SIC electronic structure obtained from the converged FOD configuration 2. Again, we check that the electronic structure does not change with further FOD relaxations or with further decrease of the dipole moment.

\begin{table}[h!]
\caption{Converged SCF total energies $E$ and maximum final force components $F_{\rm{max}}$ after the 6 frozen-density loop cycles, and energy differences $\Delta E$ between the 6-th and 5-th frozen-density loops for configuration 1-4 (labeled in Fig.~\ref{fig:FOD-Q1}) of the \textbf{Q-1} molecule. Each frozen-density loop consists of many FOD updates.}
\begin{ruledtabular}
\begin{tabular}{ccccc}
FOD  & $E$ (Ha) & $F_{\rm{max}}$   & $\Delta E$  \\
Conf &          & (mHa/$a_{\rm B}$)& (mHa)       \\ \hline
1 & -3703.29260 & 0.35 & -0.75  \\
2 & -3703.29262 & 0.34 & -0.89  \\
3 & -3703.29231 & 0.28 & -2.10  \\
4 & -3703.29250 & 0.38 & -1.04
\end{tabular}
\end{ruledtabular}
\label{tab:Q1-energy}
\end{table}


\begin{figure}[h!]
\includegraphics[width=0.9\linewidth]{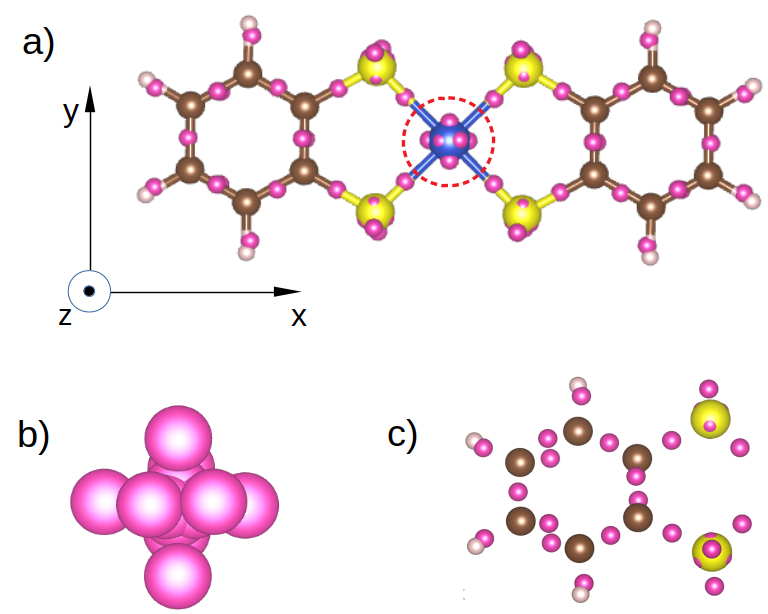}
\caption{(a) Converged spin-unpolarized FOD positions from configuration 2 for the \textbf{Q-1} molecule. (b) Zoom-in of the FODs near the Cu atom and (c) zoom-in of the FODs near the ligands.}
\label{fig:Q1-FOD}
\end{figure}

\subsection{FLO-SIC calculated electronic structure}

\begin{table}[h!]
\caption{Contributions of the Cu $d$, all S $p$, and all C $p$ orbitals to the HOMO and LUMO of the \textbf{Q-1} (charge $Q=-1$, $S=0$) molecule calculated using the FLO-SIC method compared to our SIC-free PBE-GGA calculation as well as two B3LYP results \cite{Maiti2014,Ray2005}.
In the former B3LYP calculation, the orbital decomposition was not quantified, while the latter B3LYP result is for a monoanionic Cu-based molecule with similar (slightly different) ligands, [Cu(III)(C$_{14}$H$_{20}$S$_2$)$_2$]$^{1-}$,~\cite{Ray2005}, where the HOMO and LUMO energies
were not reported. See Fig.~\ref{fig:Q1-orbs}) for the HOMO and LUMO from the FLO-SIC and PBE-GGA calculations.}
\label{tab:homo-Q1}
\begin{ruledtabular}
\begin{tabular}{cccrrr}
Method &  Level  &  Energy(eV) & Cu $d$ & S $p$ &  C $p$ \\ \hline
FLO-SIC & HOMO & -5.90 & 1.6\% &67\% &31\% \\
        & LUMO & -1.22 & 68\% &27\% &0.8\% \\  \hline
PBE-GGA & HOMO & -1.29 & 15\% &60\% &24\% \\
        & LUMO & -0.75 & 43\% &51\% &1.7\% \\ \hline
 B3LYP \cite{Maiti2014} & HOMO & $\sim$-1.5 & - & - & - \\
        & LUMO & $\sim$-0.5 & - & - & - \\ \hline
 B3LYP \cite{Ray2005} & HOMO & - & 11\% & 58\% & 27\% \\
        & LUMO & -    & 33\% & 58\% & 4\%
\end{tabular}
\end{ruledtabular}
\end{table}

Table~\ref{tab:homo-Q1} shows the energies of the HOMO and LUMO of the \textbf{Q-1} molecule calculated from the FLO-SIC method. We find that the HOMO energy is $-5.90$~eV, which is 3.91~eV lower than the HOMO energy of the \textbf{Q-2} molecule. The SIC shifts the HOMO (LUMO) level downward by 4.61 (0.67) eV in comparison to the SIC-free DFT results, such that it increases the HOMO-LUMO gap to 4.68~eV.

\begin{figure}[th!]
\centering
\includegraphics[width=0.5\textwidth]{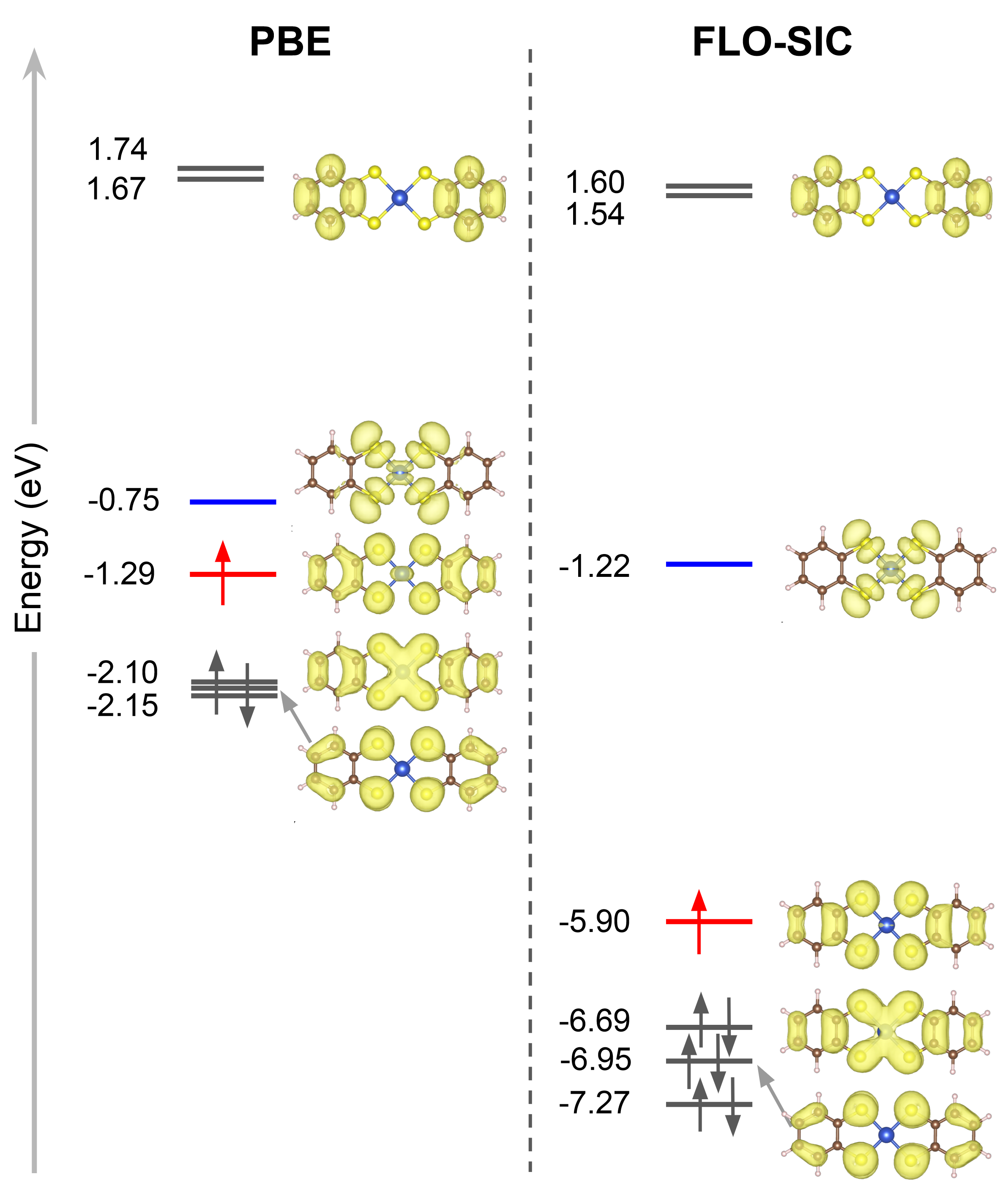}
\caption{Energy levels and corresponding canonical orbitals of the \textbf{Q-1} molecule from the FLO-SIC and SIC-free PBE-GGA calculations}
\label{fig:Q1-orbs}
\end{figure}

\begin{figure*}[th!]
\centering
\includegraphics[width=0.75\linewidth]{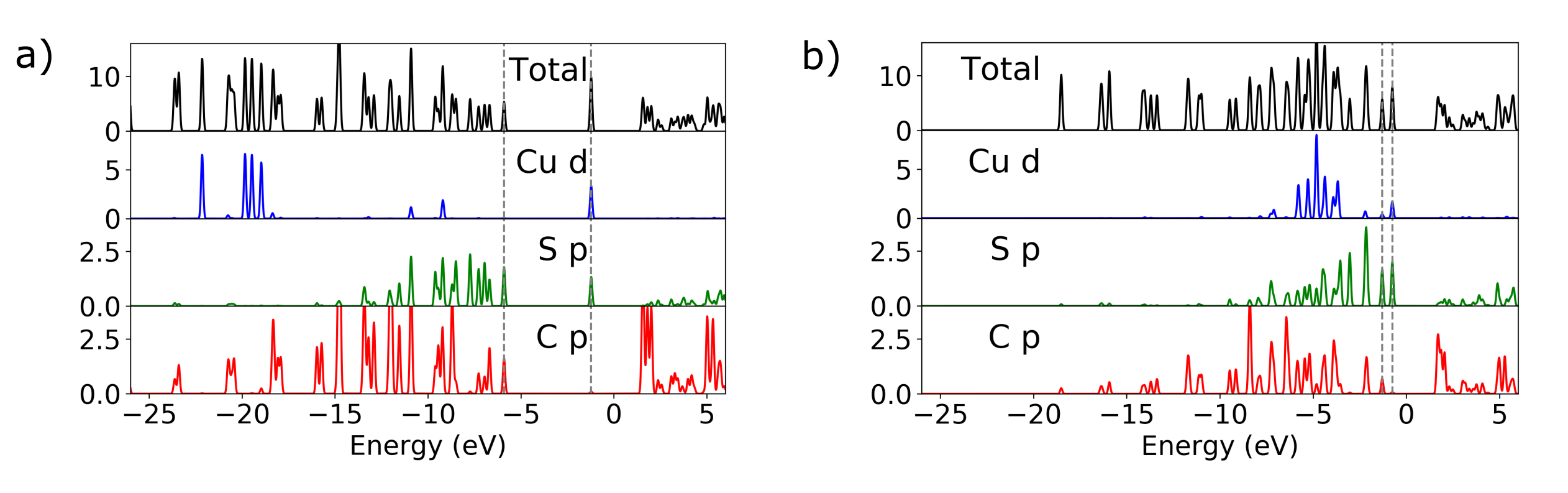}
\caption{(a) FLO-SIC and (b) SIC-free DFT calculated densities of states (DOS) projected onto the Cu $d$, S $p$, and C $p$ orbitals of
the \textbf{Q-1} molecule where the left and right vertical dashed lines indicate the HOMO and LUMO levels. The FLO-SIC HOMO-LUMO gap is 4.68 eV,
while the SIC-free DFT HOMO-LUMO gap is 0.54 eV.}
\label{fig:Q1-dos}
\end{figure*}

Figure~\ref{fig:Q1-orbs} shows the HOMO and LUMO and a few other {\it canonical} orbitals of the \textbf{Q-1} molecule calculated from the FLO-SIC method compared to the SIC-free PBE-GGA orbitals. The overall orbital shapes do not change much upon the FLO-SIC. The FLO-SIC calculation shows that the major contributions to the HOMO originate from the S $p_z$ and C $p_z$ orbitals at a ratio of almost 2:1 with a very small contribution from the Cu $d$ orbitals, while a majority contribution to the LUMO arises from the Cu $d_{xy}$ and S $p_{x,y}$ orbitals with an approximate ratio of 2:1 (see Table~\ref{tab:homo-Q1}). These characteristics somewhat differ from those of the SIC-free PBE-GGA results where the HOMO carries a significant contribution from the Cu $d_{xz}$ orbitals and the LUMO originates from almost equal contributions of Cu $d$ and S $p$ orbitals. The canonical 
orbitals away from the HOMO and LUMO from the FLO-SIC method are very similar to those from the SIC-free PBE-GGA (Fig.~\ref{fig:Q1-orbs}).


We plot the DOS projected onto Cu $d$, S $p$, and C $p$ orbitals obtained using the FLO-SIC method and the SIC-free DFT (see Fig.~\ref{fig:Q1-dos}). 
We find that the SIC lowers the occupied S $p$ and C $p$ orbitals by about 5 eV and most of the Cu 3$d$ orbitals by 14-15~eV, which is similar to the case of \textbf{Q-2}. Interestingly, for the occupied Cu $d$ orbitals hybridized with the ligands, the energy shift by the SIC is smaller, 9 eV. For \textbf{Q-1}, the HOMO character obtained from the FLO-SIC is similar to that from the SIC-free DFT because a much smaller contribution of the
Cu $d$ orbitals to the HOMO.

\subsection{Comparison with previous work} \label{sec:comp-Q1}

We now compare our FLO-SIC calculated results with the previous experimental and theoretical work~\cite{Maiti2014,Ray2005} for \textbf{Q-1}.
The characteristics and energies of the FLO-SIC HOMO and LUMO are quite different from the B3LYP results \cite{Maiti2014} (see Table~\ref{tab:homo-Q1}) which are closer to the SIC-free PBE-GGA calculation. Interestingly, the B3LYP HOMO-LUMO gap for \textbf{Q-1} is smaller than that for \textbf{Q-2} by about 1 eV. This feature is similar to the FLO-SIC result, although the gap difference is larger in the FLO-SIC case ($\sim$1.66 eV). The experimental UV-optical absorption spectra \cite{Maiti2014} show a maximum peak at 398~nm (3.1~eV) for \textbf{Q-1} and a peak at 348~nm (3.6~eV) and possible peaks below 270~nm (4.6~eV) for \textbf{Q-2}. Although the optical gap is not equivalent to the HOMO-LUMO gap \cite{Bredas2014}, it does provide some indication of the gap size. The difference in the observed gaps is thus consistent with the FLO-SIC HOMO-LUMO gap being smaller in \textbf{Q-1}.
Our results are also compared with the monoanionic Cu-based molecule with similar (slightly different) ligands such as [Cu(III)(C$_{14}$H$_{20}$S$_2$)$_2$]$^{1-}$ \cite{Ray2005}. A previous B3LYP DFT calculation \cite{Ray2005} on this monoanionic molecule with the scalar relativistic zero-order regular approximation (ZORA) showed that the HOMO consists of Cu $d_{xz}$ (11\%), S $p_z$ (58\%), and C $p_z$ (27\%), while the LUMO consists of Cu $d_{xy}$ (33\%), S $p_z$ (58\%), and C $p_z$ (4\%). Despite slightly different ligands, this orbital decomposition is consistent with our SIC-free PBE-GGA result (see Table~\ref{tab:homo-Q1}).

\section{Conclusions and Outlook}\label{sec:summary}

We have investigated the dianionic and monoanionic states of the mononuclear Cu-based molecule which has potential applications for quantum information science, using the FLO-SIC method. Starting with the SIC-free DFT-converged electron density and with multiple initial sets of FODs within the frozen-density loop approach, we determined optimal FLOs with which the SIC energy was obtained self-consistently. Although different initial sets 
of FODs converged to somewhat different final FOD positions (core and valence FODs), the electronic and magnetic properties obtained from the multiple final converged FODs are very similar to one another within our numerical accuracy. This may suggest an importance of the starting electron density 
in the FLO-SIC calculations.

As expected, the SIC potential in the FLO-SIC method remedies the severe delocalization of electric charge and spin density caused by the SIE. The FLO-SIC calculations show that in the dianionic case, about 67\% of the spin density arises from the Cu $d$ orbitals and about 33\% of the spin density is from the S $p$ orbitals. This feature quantitatively agrees with the multireference result \cite{Fataftah2019} and our multiconfigurational calculation, while the SIC-free PBE-GGA and B3LYP functionals severely underestimate the spin density of the Cu $d$ orbitals. The trend that we 
found needs to be checked for other transition-metal based molecules with covalent or ionic metal-ligand bonding. It is known that discrepancy between experimental and DFT exchange coupling constants in 3$d$ transition-metal systems, especially overestimation (without sign flip), may be mainly attributed to highly delocalized spin density of the transition-metal $d$ orbitals \cite{Rudra2006}. Considering this, our results suggest that the FLO-SIC method may be overall effective in accurately describing the magnetic exchange coupling between 3$d$ transition metal centers within multinuclear magnetic systems \cite{Joshi2018,Jackson2019}.

The SIC potential lowers the energies of the Cu $d$ orbitals by a large amount, which ended up with large changes of the HOMO character compared to the SIC-free DFT result. Whether the FLO-SIC HOMO character is adequate or not requires further studies using more advanced FLO-SIC methods. Utilization of complex orbitals in FLO-SIC calculations \cite{Klupfel2011,Lehtola2016} may improve the FOD optimization process and provide a better description of systems involving transition metals. It is shown that local scaling SIC appears important in preventing overcorrection of some properties such as polarizabilities and atomization energies \cite{Romero2021}. However, there has been so far no studies of the local scaling SIC effect on the characteristics of the HOMO, which would be also interesting to investigate.

\begin{acknowledgements}
This work was funded by the Department of Energy Basic Energy Sciences grant numbers DE-SC0019033 and DE-SC0018331. The computational support was provided by the Virginia Tech Advanced Research Computing and the Extreme Science and Engineering Discovery Environment (XSEDE) under Project number
DMR060009N, which is supported by the National Science Foundation Grant number ACI-1548562. We are grateful to Mark Pederson, Tunna Baruah, and Rajendra Zope for their extensive help in using and troubleshooting the FLOSIC code. We are also grateful to Kai Trepte for the discussion and for
providing us with the frozen-density loop functionality and the fodMC code, and we thank Aleksander Wysocki for helping us set up the CASSCF calculation.
\end{acknowledgements}

\section*{Data Availability}

The data that support the findings of this study are available within the article and the Appendices.


\appendix

\section{Molecular Geometries} \label{app:geo}

\subsection{Dianionic molecule (\textbf{Q-2})}

\begin{table}[ht!]
\caption{Geometry of the dianionic Cu-based molecule in units of $a_{\rm B}$.}
\begin{ruledtabular}
\begin{tabular}{l r r r}
Species & $x$ & $y$ & $z$ \\ \hline
Cu & 0.0000 & 0.0000 & 0.0000 \\
S & -3.0951 & -3.0344 & 0.0000 \\
S & -3.0190 & 3.0345 & 0.0000 \\
S & 3.0952 & 3.0345 & 0.0000 \\
S & 3.0191 & -3.0344 & 0.0000 \\
C & -5.9033 & -1.2856 & -0.1936 \\
C & -5.8796 & 1.3796 & -0.1871 \\
C & -10.4423 & 1.4295 & -0.5952 \\
C & -8.2349 & -2.5282 & -0.3794 \\
C & -8.1540 & 2.6950 & -0.3968 \\
C & -10.4902 & -1.1919 & -0.5795 \\
C & 5.9034 & 1.2857 & 0.1936 \\
C & 5.8797 & -1.3794 & 0.1872 \\
C & 10.4424 & -1.4294 & 0.5952 \\
C & 8.2350 & 2.5284 & 0.3794 \\
C & 8.1540 & -2.6948 & 0.3968 \\
C & 10.4903 & 1.1920 & 0.5795 \\
H & -12.1979 & 2.4933 & -0.7648 \\
H & -8.2819 & -4.5875 & -0.3674 \\
H & -8.1318 & 4.7546 & -0.4054 \\
H & -12.2829 & -2.1959 & -0.7228 \\
H & 12.1981 & -2.4933 & 0.7647 \\
H & 8.2820 & 4.5877 & 0.3675 \\
H & 8.1319 & -4.7545 & 0.4054 \\
H & 12.2830 & 2.1960 & 0.7228
\end{tabular}
\end{ruledtabular}
\label{tab:Q2-geo-app}
\end{table}


\subsection{Monoanionic molecule (\textbf{Q-1})}

\begin{table}[ht!]
\caption{Optimized geometry of the monoanionic Cu-based molecule in units of $a_{\rm B}$.}
\begin{ruledtabular}
\begin{tabular}{l r r r}
Species & $x$ & $y$ & $z$ \\ \hline
Cu & 0.0000 & 0.0000 & 0.0000 \\
S & -2.9880 & -2.9518 & -0.0110 \\
S & -2.9198 & 3.0322 & -0.0118 \\
S & 2.9882 & 2.9518 & 0.0110 \\
S & 2.9199 & -3.0323 & 0.0119 \\
C & -5.8516 & -1.2592 & -0.2332 \\
C & -5.8220 & 1.4049 & -0.2436 \\
C & -10.4096 & 1.4425 & -0.5731 \\
C & -8.1679 & -2.5462 & -0.3861 \\
C & -8.1134 & 2.7386 & -0.4114 \\
C & -10.4360 & -1.2061 & -0.5607 \\
C & 5.8517 & 1.2592 & 0.2332 \\
C & 5.8221 & -1.4050 & 0.2436 \\
C & 10.4096 & -1.4425 & 0.5730 \\
C & 8.1680 & 2.5461 & 0.3861 \\
C & 8.1136 & -2.7387 & 0.4113 \\
C & 10.4361 & 1.2060 & 0.5607 \\
H & -12.1754 & 2.5014 & -0.7241 \\
H & -8.1697 & -4.6106 & -0.3740 \\
H & -8.0702 & 4.8022 & -0.4181 \\
H & -12.2161 & -2.2373 & -0.7134 \\
H & 12.1755 & -2.5014 & 0.7241 \\
H & 8.1698 & 4.6106 & 0.3740 \\
H & 8.0703 & -4.8023 & 0.4181 \\
H & 12.2161 & 2.2373 & 0.7134
\end{tabular}
\end{ruledtabular}
\label{tab:Q1-geo-app}
\end{table}

\section{CASSCF and UHF calculations for \textbf{Q-2}}\label{app:caspt2}

\begin{figure}[h!]
    \centering
    \includegraphics[width=1.0\linewidth]{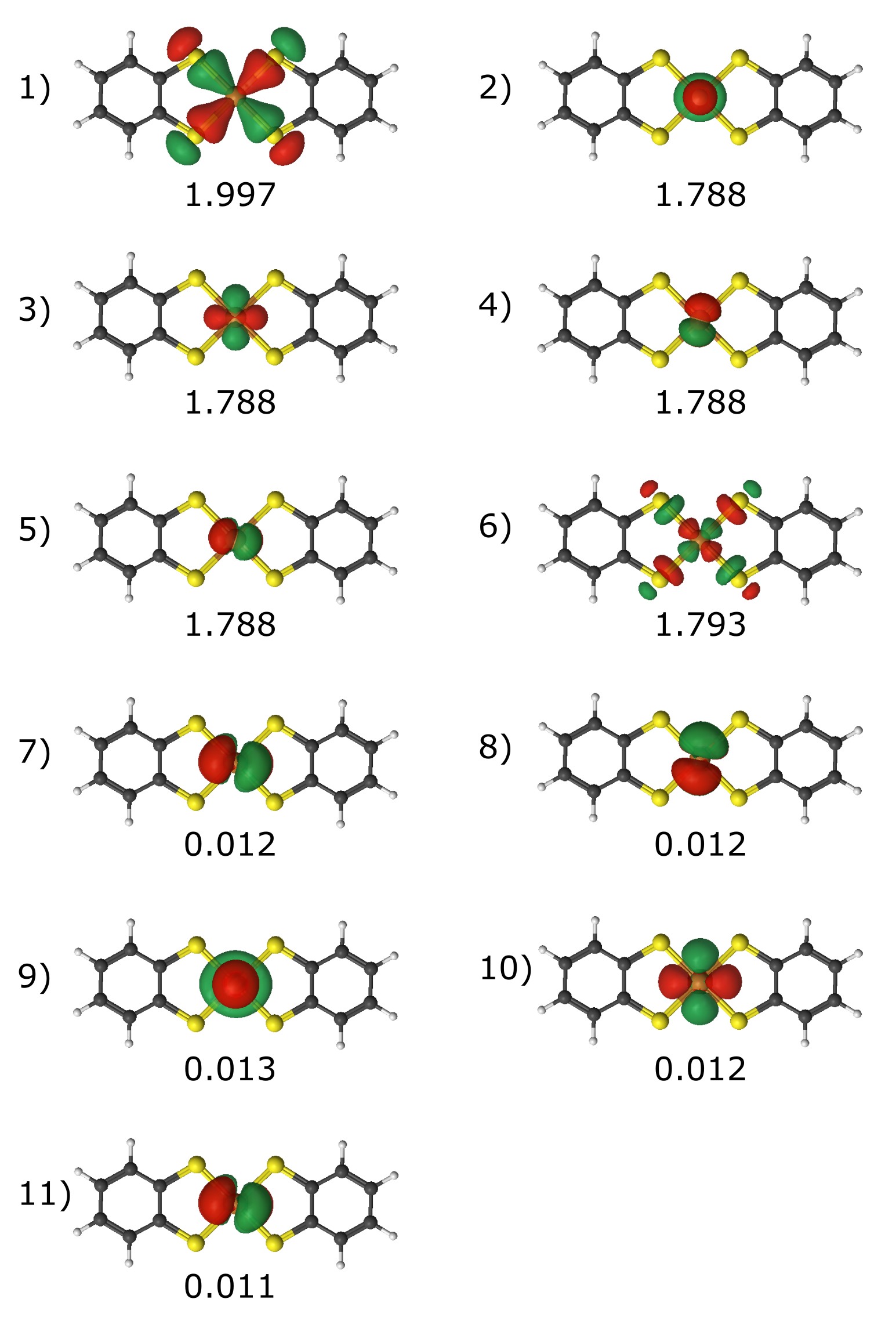}
    \caption{Molecular orbitals (MO) and state-averaged occupation numbers in the active space for our CASSCF(11,11) calculation of the \textbf{Q-2} molecule. The ground-state configuration consists of doubly occupied MO 1-5, singly occupied MO 6, and empty MO 7-11.}
    \label{fig:casorbs}
\end{figure}

We perform CASSCF calculations of the \textbf{Q-2} molecule in the doublet state ($S=1/2$) with the active space consisting of 11 electrons and 11 orbitals using {\it ab-initio} code MOLCAS version 8.2 \cite{MOLCAS}. The 11 active orbitals comprise five Cu 3$d$ orbitals, five Cu 4$d$ (or *d-shell) orbitals, and one Cu-S bonding orbital. Among the five Cu 3$d$ orbitals, one Cu-S antibonding orbital is included. The total number of electrons in the active space is counted considering the doubly occupied Cu-S bonding orbital (Cu 3$d_{xy}$ and S 3$p_{x,y}$) and nine electrons from the five Cu 3$d$. The 11 active molecular orbitals with their state-averaged occupation numbers are shown in Fig.~\ref{fig:casorbs}. Scalar relativistic effects are included based on the Douglas-Kroll-Hess Hamiltonian~\cite{Douglas1974,Hess1986} using relativistically contracted atomic natural orbital (ANO-RCC) basis sets \cite{Widmark1990,Roos2004}. For all elements, we use the valence double-$\zeta$ quality (ANO-RCC-VDZ) basis sets provided in MOLCAS code. Considering the five possibilities of creating a hole in the 3$d$ orbitals, we carried out the state average over 5 roots. From the CASSCF(11,11) calculations, we find that the ground state (lowest-energy root) has a configuration of the doubly occupied Cu-S bonding orbital (MO 1: Cu 3$d_{xy}$ and S 3$p_{x,y}$), the four doubly occupied Cu 3$d$ orbitals (MO 2-5: $d_{z^2}$, $d_{x^2-y^2}$, $d_{yz}$, $d_{xz}$), the singly occupied Cu-S antibonding orbital (MO 6: Cu 3$d_{xy}$ and S 3$p_{x,y}$), and empty Cu 4$d$ orbitals (MO 7-11).

Unrestricted Hartree-Fock (UHF) calculations are performed for the \textbf{Q-2} molecule in the doublet state ($S=1/2$) using MOLCAS version 8.2 \cite{MOLCAS}. We use ANO-RCC-VDZ basis sets for all the atoms.




\section{Converged FOD Geometries}

\subsection{Dianionic molecule: FOD Configuration 4} \label{app:Q2-FODs}


\begin{center}
\begin{longtable}{|c|c|c|c|}
\caption{Converged geometry of spin-up FODs\\
(configuration 4) for \textbf{Q-2} in units of $a_{\rm B}$.}\\
\hline
\multicolumn{4}{|c|}{Spin UP} \\ \hline
Orb. Type &$x$ &$y$ &$z$ \\ \hline
\endfirsthead

\hline
\multicolumn{4}{|c|}{Continuation of Table \ref{tab:Q2-FOD-up}} \\
\hline
Orb. Type & $x$  & $y$ & $z$ \\ \hline
\endhead

\hline \multicolumn{4}{|r|}{{Continued on the next column}} \\ \hline
\endfoot

\hline \hline
\endlastfoot

Cu 1s &0.000000 &0.000000 &0.000000 \\ \hline
\multirow{4}{*}{Cu 2sp} &0.140437 &-0.006153 &0.121301 \\ \cline{2-4}
&-0.147459 &0.008121 &0.111994 \\ \cline{2-4}
&0.011429 &0.166278 &-0.123538 \\ \cline{2-4}
&-0.004199 &-0.168263 &-0.121273 \\ \hline
\multirow{9}{*}{Cu 3spd} &0.533382 &0.563552 &0.363919 \\ \cline{2-4}
&-0.548512 &-0.557042 &0.344983 \\ \cline{2-4}
&-0.500887 &0.606451 &0.338860 \\ \cline{2-4}
&0.483260 &-0.600871 &0.370400 \\ \cline{2-4}
&0.721904 &-0.035822 &-0.585479 \\ \cline{2-4}
&-0.694333 &0.026664 &-0.625292 \\ \cline{2-4}
&0.027308 &0.407041 &-0.382308 \\ \cline{2-4}
&-0.008506 &-0.412614 &-0.377110 \\ \cline{2-4}
&-0.015814 &0.004469 &0.728797 \\ \hline
\multirow{4}{*}{Cu-S bonds} &-2.111473 &-1.842142 &0.068512 \\ \cline{2-4}
&-2.076545 &1.810656 &0.109787 \\ \cline{2-4}
&2.099305 &1.864911 &0.265062 \\ \cline{2-4}
&2.068580 &-1.828072 &0.242089 \\ \hline
\multirow{4}{*}{Ligands} &-3.095100 &-3.034400 &0.000000 \\ \cline{2-4}
&-3.101429 &-2.765981 &-0.391111 \\ \cline{2-4}
&-2.723407 &-3.305934 &0.045615 \\ \cline{2-4}
&-3.429507 &-3.361022 &0.015010 \\ \cline{2-4}
&-3.128710 &-2.693342 &0.330667 \\ \cline{2-4}
&-4.562630 &-2.068162 &-0.111434 \\ \cline{2-4}
&-3.208291 &-3.910813 &-1.368258 \\ \cline{2-4}
&-3.322393 &-3.873760 &1.378195 \\ \cline{2-4}
&-3.019000 &3.034500 &0.000000 \\ \cline{2-4}
&-3.056048 &2.695316 &0.332578 \\ \cline{2-4}
&-2.644963 &3.302568 &0.042083 \\ \cline{2-4}
&-3.025741 &2.766809 &-0.391055 \\ \cline{2-4}
&-3.352441 &3.361822 &0.017076 \\ \cline{2-4}
&-4.515427 &2.118259 &-0.105528 \\ \cline{2-4}
&-3.208008 &3.914183 &1.359502 \\ \cline{2-4}
&-3.085594 &3.878898 &-1.394130 \\ \cline{2-4}
&-5.903300 &-1.285600 &-0.193600 \\ \cline{2-4}
&-5.834331 &0.047515 &-1.226000 \\ \cline{2-4}
&-6.046157 &0.049012 &0.975923 \\ \cline{2-4}
&-7.017290 &-2.002031 &-0.273916 \\ \cline{2-4}
&-5.879600 &1.379600 &-0.187100 \\ \cline{2-4}
&-6.963985 &2.128132 &-0.275508 \\ \cline{2-4}
&-10.442300 &1.429500 &-0.595200 \\ \cline{2-4}
&-9.253654 &1.997210 &-1.351158 \\ \cline{2-4}
&-9.399692 &1.987056 &0.359386 \\ \cline{2-4}
&-10.613748 &0.121326 &-0.590620 \\ \cline{2-4}
&-11.901501 &2.120958 &-0.748147 \\ \cline{2-4}
&-8.234900 &-2.528200 &-0.379400 \\ \cline{2-4}
&-9.324935 &-1.804501 &-1.340106 \\ \cline{2-4}
&-9.455418 &-1.771954 &0.364704 \\ \cline{2-4}
&-8.109587 &-4.142157 &-0.371431 \\ \cline{2-4}
&-8.154000 &2.695000 &-0.396800 \\ \cline{2-4}
&-7.974475 &4.304984 &-0.405633 \\ \cline{2-4}
&-10.490200 &-1.191900 &-0.579500 \\ \cline{2-4}
&-11.975389 &-1.834973 &-0.715014 \\ \cline{2-4}
&3.095200 &3.034500 &0.000000 \\ \cline{2-4}
&3.118318 &2.776295 &0.397660 \\ \cline{2-4}
&2.721877 &3.298697 &-0.061242 \\ \cline{2-4}
&3.431148 &3.358185 &-0.005696 \\ \cline{2-4}
&3.112116 &2.694790 &-0.327862 \\ \cline{2-4}
&4.567332 &2.067120 &0.092021 \\ \cline{2-4}
&3.239123 &4.127233 &1.197780 \\ \cline{2-4}
&3.281845 &3.624508 &-1.524598 \\ \cline{2-4}
&3.019100 &-3.034400 &0.000000 \\ \cline{2-4}
&3.036860 &-2.694898 &-0.329390 \\ \cline{2-4}
&2.644071 &-3.296117 &-0.061708 \\ \cline{2-4}
&3.044641 &-2.776198 &0.398150 \\ \cline{2-4}
&3.354005 &-3.359786 &-0.004370 \\ \cline{2-4}
&4.517107 &-2.116384 &0.083802 \\ \cline{2-4}
&3.183481 &-3.654536 &-1.512668 \\ \cline{2-4}
&3.105700 &-4.110867 &1.215247 \\ \cline{2-4}
&5.903400 &1.285700 &0.193600 \\ \cline{2-4}
&5.855322 &-0.047856 &1.271607 \\ \cline{2-4}
&6.013191 &-0.048392 &-0.924280 \\ \cline{2-4}
&7.016180 &2.002949 &0.282524 \\ \cline{2-4}
&5.879700 &-1.379400 &0.187200 \\ \cline{2-4}
&6.963033 &-2.128481 &0.284354 \\ \cline{2-4}
&10.442400 &-1.429400 &0.595200 \\ \cline{2-4}
&9.247638 &-1.983879 &1.356095 \\ \cline{2-4}
&9.406294 &-1.999631 &-0.356898 \\ \cline{2-4}
&10.611694 &-0.121133 &0.600671 \\ \cline{2-4}
&11.901929 &-2.122664 &0.730683 \\ \cline{2-4}
&8.235000 &2.528400 &0.379400 \\ \cline{2-4}
&9.318521 &1.791676 &1.345438 \\ \cline{2-4}
&9.462484 &1.784652 &-0.362645 \\ \cline{2-4}
&8.111172 &4.141376 &0.353021 \\ \cline{2-4}
&8.154000 &-2.694800 &0.396800 \\ \cline{2-4}
&7.976123 &-4.304237 &0.387162 \\ \cline{2-4}
&10.490300 &1.192000 &0.579500 \\ \cline{2-4}
&11.975580 &1.836666 &0.697490

\label{tab:Q2-FOD-up}

\end{longtable}
\end{center}

\begin{center}
\begin{longtable}{|c|c|c|c|}
\caption{Converged geometry of spin-down FODs \\
 (configuration 4) for \textbf{Q-2} in units of $a_{\rm B}$.}\\
 \hline
 \multicolumn{4}{|c|}{Spin DOWN} \\ \hline
Orb. Type & $x$ & $y$ & $z$ \\ \hline
\endfirsthead

\hline
\multicolumn{4}{|c|}{Continuation of Table \ref{tab:Q2-FOD-dn}} \\
\hline
Orb. Type & $x$  & $y$ & $z$ \\ \hline
\endhead

\hline \multicolumn{4}{|r|}{{Continued on the next column}} \\ \hline
\endfoot

\hline \hline
\endlastfoot

Cu 1s &0.000000 &0.000000 &0.000000 \\ \hline
\multirow{4}{*}{Cu 2sp} &0.130298 &-0.027391 &-0.156584 \\ \cline{2-4}
&-0.174223 &-0.026294 &-0.059652 \\ \cline{2-4}
&0.016902 &0.176076 &0.058717 \\ \cline{2-4}
&0.031878 &-0.127228 &0.157324 \\ \hline
\multirow{8}{*}{Cu 3spd} &0.840120 &0.011401 &0.249996 \\ \cline{2-4}
&-0.755827 &-0.015060 &0.262829 \\ \cline{2-4}
&-0.034422 &0.415893 &0.654691 \\ \cline{2-4}
&0.011549 &-0.310753 &0.473532 \\ \cline{2-4}
&0.310661 &-0.011339 &-0.470749 \\ \cline{2-4}
&-0.421930 &0.035834 &-0.668717 \\ \cline{2-4}
&0.012690 &0.749680 &-0.259756 \\ \cline{2-4}
&-0.003356 &-0.840355 &-0.247173 \\ \hline
\multirow{4}{*}{Cu-S bonds} &-1.944906 &-1.856133 &-0.108011 \\ \cline{2-4}
&-1.889778 &1.810019 &-0.024370 \\ \cline{2-4}
&1.968248 &1.831567 &0.075125 \\ \cline{2-4}
&1.869221 &-1.791551 &0.012073 \\ \hline
\multirow{4}{*}{Ligands} &-3.095100 &-3.034400 &0.000000 \\ \cline{2-4}
&-3.104623 &-2.792723 &-0.399787 \\ \cline{2-4}
&-2.713020 &-3.281980 &0.047461 \\ \cline{2-4}
&-3.124771 &-2.714818 &0.339812 \\ \cline{2-4}
&-3.442438 &-3.342633 &0.012001 \\ \cline{2-4}
&-3.091948 &-3.953547 &-1.367253 \\ \cline{2-4}
&-3.249423 &-3.711402 &1.507150 \\ \cline{2-4}
&-3.019000 &3.034500 &0.000000 \\ \cline{2-4}
&-3.031520 &2.795520 &-0.397043 \\ \cline{2-4}
&-2.635299 &3.275305 &0.043065 \\ \cline{2-4}
&-3.050727 &2.720721 &0.340219 \\ \cline{2-4}
&-3.362062 &3.341376 &0.013373 \\ \cline{2-4}
&-3.146003 &3.787235 &1.500560 \\ \cline{2-4}
&-3.013779 &3.960913 &-1.402001 \\ \cline{2-4}
&-5.903300 &-1.285600 &-0.193600 \\ \cline{2-4}
&-4.577437 &-1.995539 &-0.094181 \\ \cline{2-4}
&-5.879600 &1.379600 &-0.187100 \\ \cline{2-4}
&-4.545042 &2.045505 &-0.094045 \\ \cline{2-4}
&-5.696598 &0.044912 &-0.172740 \\ \cline{2-4}
&-10.442300 &1.429500 &-0.595200 \\ \cline{2-4}
&-8.234900 &-2.528200 &-0.379400 \\ \cline{2-4}
&-7.105984 &-1.905376 &-1.234125 \\ \cline{2-4}
&-7.251680 &-1.911673 &0.660927 \\ \cline{2-4}
&-8.154000 &2.695000 &-0.396800 \\ \cline{2-4}
&-7.045929 &2.032268 &-1.250852 \\ \cline{2-4}
&-7.201763 &2.052201 &0.650728 \\ \cline{2-4}
&-9.338694 &2.135280 &-0.499804 \\ \cline{2-4}
&-10.490200 &-1.191900 &-0.579500 \\ \cline{2-4}
&-10.535883 &0.126545 &0.369795 \\ \cline{2-4}
&-10.359909 &0.108835 &-1.541474 \\ \cline{2-4}
&-9.405401 &-1.931696 &-0.482550 \\ \cline{2-4}
&-11.761447 &2.344416 &-0.719292 \\ \cline{2-4}
&-8.373172 &-4.109167 &-0.374801 \\ \cline{2-4}
&-8.236908 &4.279416 &-0.407785 \\ \cline{2-4}
&-11.841502 &-2.063441 &-0.684648 \\ \cline{2-4}
&3.095200 &3.034500 &0.000000 \\ \cline{2-4}
&3.099965 &2.792682 &0.399867 \\ \cline{2-4}
&2.712265 &3.281476 &-0.043372 \\ \cline{2-4}
&3.129430 &2.716851 &-0.341432 \\ \cline{2-4}
&3.443544 &3.341366 &-0.014693 \\ \cline{2-4}
&3.092415 &3.916495 &1.391044 \\ \cline{2-4}
&3.253715 &3.761675 &-1.484140 \\ \cline{2-4}
&3.019100 &-3.034400 &0.000000 \\ \cline{2-4}
&3.028511 &-2.795451 &0.395219 \\ \cline{2-4}
&2.635777 &-3.274514 &-0.039840 \\ \cline{2-4}
&3.054223 &-2.722745 &-0.339478 \\ \cline{2-4}
&3.361099 &-3.340366 &-0.015591 \\ \cline{2-4}
&3.161725 &-3.825869 &-1.487450 \\ \cline{2-4}
&3.012624 &-3.937765 &1.421397 \\ \cline{2-4}
&5.903400 &1.285700 &0.193600 \\ \cline{2-4}
&4.569765 &1.992810 &0.093496 \\ \cline{2-4}
&5.879700 &-1.379400 &0.187200 \\ \cline{2-4}
&4.548281 &-2.047422 &0.089451 \\ \cline{2-4}
&5.696742 &-0.044162 &0.173826 \\ \cline{2-4}
&10.442400 &-1.429400 &0.595200 \\ \cline{2-4}
&8.235000 &2.528400 &0.379400 \\ \cline{2-4}
&7.103359 &1.906668 &1.225390 \\ \cline{2-4}
&7.256631 &1.910027 &-0.670244 \\ \cline{2-4}
&8.154000 &-2.694800 &0.396800 \\ \cline{2-4}
&7.043586 &-2.032815 &1.243270 \\ \cline{2-4}
&7.205103 &-2.051525 &-0.660238 \\ \cline{2-4}
&9.338475 &-2.134622 &0.499690 \\ \cline{2-4}
&10.490300 &1.192000 &0.579500 \\ \cline{2-4}
&10.534110 &-0.126374 &-0.372469 \\ \cline{2-4}
&10.362186 &-0.109057 &1.541334 \\ \cline{2-4}
&9.405378 &1.931580 &0.482034 \\ \cline{2-4}
&11.761808 &-2.343494 &0.721289 \\ \cline{2-4}
&8.372809 &4.109440 &0.375394 \\ \cline{2-4}
&8.235965 &-4.279196 &0.407798 \\ \cline{2-4}
&11.841594 &2.062818 &0.686736
\label{tab:Q2-FOD-dn}
\end{longtable}
\end{center}


\subsection{Monoanionic molecule: FOD Configuration 2} \label{app:Q1-FODs}

\begin{center}
\begin{longtable}{|c|c|c|c|}
\caption{Converged FOD geometry of configuration 2 \\
for \textbf{Q-1} in units of $a_{\rm B}$ }\\\hline

Orb. Type & $x$  & $y$ & $z$ \\ \hline
\endfirsthead

\hline
\multicolumn{4}{|c|}{Continuation of Table \ref{tab:Q1-FODs-app}} \\
\hline
Orb. Type & $x$  & $y$ & $z$ \\ \hline
\endhead

\hline \multicolumn{4}{|r|}{{Continued on the next column}} \\ \hline
\endfoot

\hline \hline
\endlastfoot

Cu  1s & 0.000000 & 0.000000 & 0.000000 \\ \hline
\multirow{4}{*}{Cu 2sp} & 0.171963 & -0.021944 & 0.068097 \\ \cline{2-4}
 & -0.132286 & -0.023862 & 0.154110 \\ \cline{2-4}
 & -0.014376 & 0.173357 & -0.067050 \\ \cline{2-4}
 & -0.027987 & -0.130181 & -0.155197 \\ \hline
\multirow{8}{*}{Cu 3spd} & 0.394996 & 0.024379 & 0.617239 \\ \cline{2-4}
 & -0.001482 & -0.823311 & 0.249773 \\ \cline{2-4}
 & -0.310798 & -0.007790 & 0.469381 \\ \cline{2-4}
 & 0.746606 & -0.016562 & -0.261254 \\ \cline{2-4}
 & 0.036822 & 0.396496 & -0.614567 \\ \cline{2-4}
 & -0.820871 & 0.023746 & -0.252955 \\ \cline{2-4}
 & 0.003839 & 0.744894 & 0.262968 \\ \cline{2-4}
 & -0.014743 & -0.308952 & -0.470619 \\ \hline
\multirow{4}{*}{Cu-S bonds} & -1.781915 & -1.691625 & -0.018311 \\ \cline{2-4}
 & -1.791331 & 1.782652 & -0.109085 \\ \cline{2-4}
 & 1.797969 & 1.710358 & 0.008900 \\ \cline{2-4}
 & 1.780486 & -1.790712 & 0.093198 \\ \hline
\multirow{4}{*}{Ligands} & -2.987407 & -2.951645 & -0.010186 \\ \cline{2-4}
 & -3.345872 & -3.242831 & 0.004111 \\ \cline{2-4}
 & -2.595029 & -3.173913 & 0.026158 \\ \cline{2-4}
 & -3.025560 & -2.657471 & 0.339888 \\ \cline{2-4}
 & -2.992640 & -2.732125 & -0.416618 \\ \cline{2-4}
 & -4.531494 & -2.019853 & -0.133993 \\ \cline{2-4}
 & -2.878634 & -3.750841 & -1.469355 \\ \cline{2-4}
 & -3.044755 & -3.598620 & 1.526392 \\ \cline{2-4}
 & -2.919038 & 3.032393 & -0.012400 \\ \cline{2-4}
 & -2.941888 & 2.746792 & -0.376666 \\ \cline{2-4}
 & -2.568019 & 3.319207 & 0.041078 \\ \cline{2-4}
 & -3.224241 & 3.386509 & 0.003995 \\ \cline{2-4}
 & -2.952933 & 2.675032 & 0.286776 \\ \cline{2-4}
 & -4.460880 & 2.143257 & -0.118220 \\ \cline{2-4}
 & -3.009789 & 3.607845 & 1.520725 \\ \cline{2-4}
 & -2.720682 & 3.775947 & -1.449289 \\ \cline{2-4}
 & 2.988705 & 2.951802 & 0.010166 \\ \cline{2-4}
 & 3.347068 & 3.240376 & -0.003442 \\ \cline{2-4}
 & 2.991387 & 2.733508 & 0.418910 \\ \cline{2-4}
 & 3.025331 & 2.658371 & -0.342131 \\ \cline{2-4}
 & 2.591924 & 3.173543 & -0.026491 \\ \cline{2-4}
 & 4.529326 & 2.017766 & 0.138384 \\ \cline{2-4}
 & 2.876767 & 3.741768 & 1.470373 \\ \cline{2-4}
 & 3.030660 & 3.589093 & -1.525161 \\ \cline{2-4}
 & 2.920530 & -3.032262 & 0.012655 \\ \cline{2-4}
 & 3.218776 & -3.389315 & 0.002669 \\ \cline{2-4}
 & 2.945682 & -2.672171 & -0.283756 \\ \cline{2-4}
 & 2.949366 & -2.744008 & 0.373499 \\ \cline{2-4}
 & 2.568700 & -3.323262 & -0.047971 \\ \cline{2-4}
 & 4.460029 & -2.132216 & 0.126484 \\ \cline{2-4}
 & 2.992370 & -3.624563 & -1.510985 \\ \cline{2-4}
 & 2.728562 & -3.793353 & 1.438725 \\ \cline{2-4}
 & -5.868791 & -1.332683 & -0.231923 \\ \cline{2-4}
 & -5.922333 & 0.079014 & 0.794537 \\ \cline{2-4}
 & -5.786270 & 0.069544 & -1.306878 \\ \cline{2-4}
 & -6.937767 & -2.009568 & -0.306224 \\ \cline{2-4}
 & -5.839102 & 1.478004 & -0.244650 \\ \cline{2-4}
 & -6.898334 & 2.169549 & -0.327058 \\ \cline{2-4}
 & -10.459506 & 1.390212 & -0.575636 \\ \cline{2-4}
 & -9.194208 & 2.012593 & -1.360434 \\ \cline{2-4}
 & -9.329891 & 2.028817 & 0.374653 \\ \cline{2-4}
 & -10.513841 & 0.119294 & -0.572553 \\ \cline{2-4}
 & -11.835054 & 2.173869 & -0.687385 \\ \cline{2-4}
 & -8.099575 & -2.565909 & -0.382148 \\ \cline{2-4}
 & -9.372209 & -1.804942 & 0.398234 \\ \cline{2-4}
 & -9.233172 & -1.806713 & -1.329643 \\ \cline{2-4}
 & -8.025888 & -4.128546 & -0.360571 \\ \cline{2-4}
 & -8.044570 & 2.756914 & -0.406001 \\ \cline{2-4}
 & -7.944740 & 4.321873 & -0.402115 \\ \cline{2-4}
 & -10.484763 & -1.152881 & -0.565010 \\ \cline{2-4}
 & -11.874154 & -1.913766 & -0.673746 \\ \cline{2-4}
 & 5.869096 & 1.332206 & 0.231920 \\ \cline{2-4}
 & 5.925368 & -0.080192 & -0.804028 \\ \cline{2-4}
 & 5.783175 & -0.070854 & 1.316307 \\ \cline{2-4}
 & 6.938917 & 2.008504 & 0.304771 \\ \cline{2-4}
 & 5.839239 & -1.478021 & 0.244374 \\ \cline{2-4}
 & 6.901224 & -2.165487 & 0.325413 \\ \cline{2-4}
 & 10.459557 & -1.390233 & 0.575542 \\ \cline{2-4}
 & 9.328372 & -2.027619 & -0.377502 \\ \cline{2-4}
 & 9.193831 & -2.013564 & 1.364061 \\ \cline{2-4}
 & 10.513699 & -0.119602 & 0.571584 \\ \cline{2-4}
 & 11.835639 & -2.173955 & 0.689013 \\ \cline{2-4}
 & 8.099563 & 2.565841 & 0.382213 \\ \cline{2-4}
 & 9.234255 & 1.807794 & 1.329987 \\ \cline{2-4}
 & 9.371656 & 1.803899 & -0.398104 \\ \cline{2-4}
 & 8.025786 & 4.128456 & 0.362188 \\ \cline{2-4}
 & 8.044535 & -2.756844 & 0.405958 \\ \cline{2-4}
 & 7.950583 & -4.325674 & 0.404086 \\ \cline{2-4}
 & 10.484870 & 1.152872 & 0.565014 \\ \cline{2-4}
 & 11.873725 & 1.914077 & 0.675239 \\ \cline{2-4}

\label{tab:Q1-FODs-app}
\end{longtable}

\end{center}

\section{Final FOD-configuration dependence of HOMO-LUMO gap and spin density for \textbf{Q-2} and \textbf{Q-1}} \label{app:FOD-comp}

\begin{table*}[hb!]
\caption{Final FOD dependence of the electronic structure such as the HOMO and LUMO energies, HOMO-LUMO gap, and Mulliken spin population on the Cu
and S atoms for \textbf{Q-2}. The final converged FOD configurations starting from FOD configuration 2-8 are used for the calculation.}
\label{tab:Q2-dep-HL}
\begin{ruledtabular}
\begin{tabular}{lrrrrrrr}
Property          &  Conf.2 & Conf.3 & Conf.4 & Conf.5 & Conf.6 & Conf.7 & Conf.8 \\ \hline
HOMO energy (eV)  & -1.99	& -1.99  & -1.99  &	-1.99  & -1.99  & -1.99	 & -1.99 \\
LUMO energy (eV)  & 4.35	& 4.35	 & 4.35	  & 4.37   & 4.38   & 4.34   &	4.34 \\
HOMO-LUMO gap (eV)& 6.34	& 6.34	 & 6.34	  & 6.36   & 6.37	& 6.33	 & 6.34 \\ \hline
Spin Cu ($\mu_B$) & 0.666	& 0.666  &	0.665 &	0.660  & 0.658  & 0.670  &	0.674 \\
Spin  S ($\mu_B$) & 0.333   & 0.333  &	0.334 &	0.337  & 0.340  & 0.326  &	0.321
\end{tabular}
\end{ruledtabular}
\end{table*}

\begin{table*}[ht!]
\caption{Final FOD dependence of the electronic structure such as the HOMO and LUMO energies and HOMO-LUMO gap for \textbf{Q-1}. The final
converged FOD configurations starting from FOD configuration 1-4 are used for the calculations.}
\label{tab:Q1-dep-HL}
\begin{ruledtabular}
\begin{tabular}{lrrrr}
Property          &  Conf.1 & Conf.2 & Conf.3 & Conf.4  \\ \hline
HOMO energy (eV)  & -5.90	& -5.90  & -5.90  &	-5.90   \\
LUMO energy (eV)  & -1.22	& -1.22  & -1.22  & -1.22   \\
HOMO-LUMO gap (eV)&  4.68	&  4.68	 &  4.68  &  4.68
\end{tabular}
\end{ruledtabular}
\end{table*}

\clearpage

\bibliography{ms.bib}
\end{document}